\documentclass[10pt]{article}  

\usepackage{graphicx}
\usepackage[utf8]{inputenc}
\usepackage[a4paper,left=2cm,right=2.cm, bottom=2.cm, top=2.0cm]{geometry}
\usepackage{amssymb}
\usepackage{amsmath}
\usepackage{bm}
\usepackage[margin=1pt,font=small,labelfont=bf]{caption}
\usepackage[dvipsnames]{xcolor}
\definecolor{citecolor}{RGB}{128,0,32}
\usepackage[unicode, pdfencoding=auto,psdextra,hyperfootnotes=false,breaklinks=true,colorlinks=true,allcolors=citecolor]{hyperref}
\usepackage{xr-hyper}
\usepackage{setspace}
\usepackage{fancyhdr}
\usepackage{titlesec}
\usepackage{algorithm}
\usepackage{algpseudocode}
\usepackage{etoolbox}

\usepackage{times}
\usepackage{tabularx}
\usepackage{datetime} 
\usepackage{lineno}
\usepackage{booktabs}
\usepackage{etoolbox}
\usepackage{subcaption}
\usepackage[normalem]{ulem}
\patchcmd{\linenumber}{\hb@xt@}{\hbox}{}{}
\newdateformat{usvardate}{\shortmonthname[\THEMONTH]. \THEDAY, \THEYEAR}

\newcommand{\lastmodified}{%
  \textcolor{gray}{\scriptsize \usvardate\today{} at \currenttime}
}

\fancypagestyle{custom}
{      
\fancyhead[L]{}
\fancyhead[C]{\textcolor{gray}{\scriptsize manuscript under review \textit{Journal of Geophysical Research}}}
\fancyhead[R]{}
\fancyfoot[C]{\textcolor{gray}{\thepage}}
\fancyfoot[R]{\lastmodified}
\fancyfoot[L]{}
}
\renewcommand\thesection{\arabic{section}}
\titlespacing\section{0pt}{12pt plus 4pt minus 2pt}{0pt plus 2pt minus 2pt}

\usepackage[
  giveninits=true,
  style=authoryear,
  backend=bibtex,
  maxcitenames=2,
  maxbibnames=99,
  hyperref=true,
  backref=false,
  sorting=nyt,
  uniquename=init,
  autolang=hyphen,
  dashed=false]{biblatex}

\AtEveryCite{\it \color{citecolor}}
\AtEveryBibitem{\clearfield{month}}
\DeclareGraphicsExtensions{.jpg}

\usepackage[breakable]{tcolorbox}
\definecolor{harsha}{RGB}{200,200,255}

\newtcbox{\inlinehl}{
  on line,
  colback=harsha!70,
  colframe=harsha!70,
  boxrule=0pt,
  arc=1mm,
  left=2pt,
  right=2pt,
  top=1pt,
  bottom=1pt,
}

\usepackage{xcolor}

\newif\ifrevisions
\revisionsfalse

\ifrevisions
  \newenvironment{revblock}{\begingroup\color{blue}}{\endgroup}
  \newcommand{\rev}[1]{\textcolor{blue}{#1}}
\else
  \newenvironment{revblock}{}{}
  \newcommand{\rev}[1]{#1}
\fi

\begin{document}
 
\pagestyle{custom}

\begin{center}
\LARGE {\bf Theoretical constraints on tidal triggering of slow earthquakes \\[12pt]}

\normalsize

Yishuo Zhou$^{1}$, 
Ankit Gupta$^{1,\dagger}$, 
Hideo Aochi$^{1,2}$, 
Alexandre Schubnel$^{1}$, 
Satoshi Ide$^{3}$,
Pierpaolo Dubernet$^1$,
Harsha S. Bhat$^{1}$

\begin{enumerate}
\small
\setlength\itemsep{-5pt}
\item {Laboratoire de G\'{e}ologie, Ecole Normale Sup\'{e}rieure, CNRS-UMR 8538, PSL Research University, Paris, France}
\item {Bureau de Recherches G\'{e}ologiques et Mini\'{e}res (BRGM), 45100 Orl\'{e}ans, France}
\item {Department of Earth and Planetary Science, The University of Tokyo, Tokyo, Japan}
\item[$\dagger$]Corresponding author: \texttt{ankit.gupta@ens.fr}
\end{enumerate}

{ \small
\textbf{Keywords: }
 Tidal triggering;
Small stress perturbations;
Rate-and-state friction;
Slow earthquakes;
Frictional properties}
\end{center}

\section*{CRediT}

\scriptsize

\begin{tabularx}{\textwidth}{rX}
\textbf{Conceptualization:} & {H. Aochi, A. Schubnel, H. S. Bhat}   \\
\textbf{Methodology:} & { A. Gupta, Y. Zhou}  \\
\textbf{Software:} & { A. Gupta, Y. Zhou, P. Dubernet} \\
\textbf{Investigation:} & { Y. Zhou, A. Gupta} \\
\textbf{Writing -- original draft:} & { Y. Zhou} \\
\textbf{Writing -- review \& editing:} & { H. Aochi, A. Schubnel, H. S. Bhat, S. Ide, A. Gupta, Y. Zhou}  \\
\textbf{Supervision:} & { H. Aochi, A. Schubnel, H. S. Bhat} \\
\textbf{Funding acquisition:} & { A. Schubnel, H. Aochi, H. S. Bhat} 
\end{tabularx}

\small 
\section*{Abstract}
Tidal stress is a globally acting  perturbation driven primarily by the gravitational forces of the Moon and the Sun. Understanding how tidal stresses can trigger seismic events is essential for constraining tectonic environments that are sensitive to small stress perturbations. Here, employing a spring–block model with rate-and-state friction, we investigate tidal triggering on velocity-weakening stable sliding faults with stiffness slightly exceeding the critical stiffness. We first apply  a step and a boxcar with finite duration normal stress perturbation to demonstrate a resonance-like amplification of slip velocity for specific boxcar durations. Next, we perform nondimensional analyses and numerical simulations with harmonic perturbations to identify the key parameters controlling tidal triggering and their admissible ranges. Triggered slip events are further characterized using physically observable quantities, including radiation efficiency and tidal phase.
Our results show that even small stress perturbations can trigger periodic as well as temporally complex slip events on stable sliding faults. The triggering behavior is primarily controlled by the normalized perturbation period and the normalized perturbation amplitude. An increase in the normalized period shifts event timing from the peak of tidal stress toward the peak of stress rate, whereas increasing the normalized amplitude promotes a transition from slow to fast events. This framework helps explain the period-dependent sensitivity and the observed phase preference between tidal stress and maximum slip velocity. Comparison between observed and model-predicted tidal correlation patterns may therefore help constrain the instantaneous frictional strength of the interface, as well as the characteristic slip distance for frictional weakening.

\section*{Plain Language Summary}
Earth's crust experiences tiny but continuous stress changes caused by the gravitational forces of the Moon and Sun, known as tidal stresses. Although these stresses are very small, typically of the order of a few kilopascals and comparable to the pressure from a gentle hand press, they have been observed to trigger slow earthquakes on some faults. This raises an important question about the physical conditions under which faults become sensitive to such weak, periodic forces. Here, we show that even very weak tidal stresses can induce faults to slip through a process called resonance, much like pushing a swing at the right rhythm makes it move higher. We find that when the period of tidal stress perturbations matches the natural response timescale of a fault, the fault becomes significantly more sensitive to triggering. Using nondimensional analysis and numerical simulations, we identify the specific combinations of fault properties and tidal characteristics that lead to slip events. Our work helps explain why some faults are unusually sensitive to tiny tidal forces and provides a way to infer fault properties from observations of tidal triggered slow earthquakes.

\section*{Keypoints}

\begin{itemize}
\begin{revblock}
\item Small harmonic stress perturbations can trigger slow and fast events through slip velocity amplification via fault resonance.
\item Triggering occurs only within specific perturbation regimes, with a transition from stress-controlled to stress-rate-controlled modulation.
\item This resonance framework can help constrain both the instantaneous frictional strength and characteristic slip distance.
\end{revblock}
\end{itemize}

\normalsize
\section{Introduction}
Slow earthquakes are a class of fault slip phenomena that release tectonic stress over timescales much longer than ordinary earthquakes, but shorter than long-term stable sliding. Over the past two and a half decades, slow earthquakes have been detected widely in many subduction zones around the Pacific Rim \parencite{schwartz2007slow,peng2010integrated}. Since they are predominantly observed both updip and downdip of megathrust seismogenic zones \parencite{obara2016connecting,nishikawa2023review}, slow earthquakes provide valuable insights into stress accumulation and release along plate interfaces and are therefore highly relevant for assessing the rupture potential and spatial extent of future megathrust earthquakes \parencite{obara2025slow}.

Slow earthquakes can be broadly classified into five types, which are commonly grouped according to their detection methods as either seismic or geodetic. 
Seismically detected events, identified through ground shaking recorded by seismometers, include low-frequency earthquakes (LFEs), tectonic tremors (hereafter referred to as tremors), and very low-frequency earthquakes (VLFEs). 
Geodetically detected events, identified through crustal deformation measured by GNSS, strainmeters, or tiltmeters, include short-term and long-term slow slip events (SSEs). 
Among these phenomena, tremor is characterized by weak seismic signals lacking clear $\rm P$- and $\rm S$-wave arrivals \parencite{obara2002nonvolcanic}, with dominant frequencies in the $1$--$10$~$\rm Hz$ range \parencite{shelly2007non,obara2020characteristic}. A key feature of tremor is its frequent spatial and temporal association with short-term SSEs, a coupled phenomenon now widely known as episodic tremor and slip (ETS) \parencite{rogers2003episodic,obara2004episodic}. Because tremor is the most frequently detected signal, it is often used as a proxy for studying ETS \parencite{bartlow2011space,wech2025rupture}.

Another important feature of slow earthquakes, particularly tremors, is their high correlation with tidal stresses \parencite{ide2014controls,yabe2015tidal,ide2015thrust,hirose2025tidal}. 
Tidal stress acting on the Earth arises from gravitational forces by the Moon and the Sun, manifested through both solid Earth tides and ocean tidal loading, and produces well-characterized harmonic stress perturbations with dominant periods of approximately 12~hours and amplitudes on the order of kilopascals ($\sim$~$\rm kPa$). 
Early observations in southwest Japan and Cascadia showed that tremor occurrences often peak at intervals of about $12$ and $24$~hours, reflecting tidal periodicity \parencite{shelly2007complex,rubinstein2008tidal,nakata2008non}. 
Motivated by these findings, subsequent studies have employed a variety of statistical approaches to quantify the relationship between tides and tremor occurrence, commonly by analyzing the distribution of tremors and LFEs with respect to tidal phase \parencite{thomas2012tidal,royer2015tidal,van2016fortnightly}.

Numerous rate-and-state friction (RSF) models have been proposed to explain observed tidal correlations and relate them to underlying fault properties. Existing RSF models of tidal correlation can be broadly divided into two categories based on the slip regime of the patch in the absence of external perturbations. In the first category, the patch is already in an unstable sliding regime without perturbations and would generate events spontaneously even in the absence of external perturbation; in this case, harmonic stress perturbation primarily modulates the naturally recurring events. In the second category, the patch undergoes stable sliding without perturbations, such that small harmonic stress perturbations can directly trigger events or amplify the slip velocity on an otherwise stable sliding patch. In the RSF spring-block framework, the fault exhibits velocity weakening (VW) behavior when the friction parameter $a/b<1$, which favors stick-slip behaviors for stiffness below the critical value $k<k_c$ but stable sliding for stiffness above it $k>k_c$. By contrast, when $a/b>1$, velocity strengthening (VS) behavior promotes stable sliding \parencite{ruina1983}. 

In the first category, \textcite{dieterich1994constitutive} developed the classical seismicity-rate framework, which relates the stressing history to the seismicity rate based on a population of independent VW spring-block systems undergoing stick-slip behaviors ($k<k_c$), assuming a no-healing approximation during nucleation. Within this framework, the modulation response is primarily controlled by the seismicity relaxation timescale $t_a$, which characterizes the timescale over which seismicity relaxes to the background level. \textcite[][Appendix~A2.2]{rubin2005earthquake} further showed that the no-healing approximation during nucleation in the spring-block model is primarily valid when $k \ll k_c$, i.e., the source patch size is much larger than its nucleation length.

Building on this framework, laboratory observations and analytical studies \parencite{beeler2003earthquakes,heimisson2020analytical} showed that short-period perturbation tends to produce modulation approximately in phase with stress, whereas long-period perturbation becomes more closely correlated with stressing rate. This seismicity-rate framework has been applied to interpret observed tidal and seasonal seismicity modulation and constrain the friction parameter $a\sigma$, seismicity relaxation timescale $t_a$ \parencite{lu2025exploring,udelllopez2026,sirorattanakul2026seismic}. Subsequent studies further incorporated stress interactions between patches \parencite{ziv2003implications,heimissonConstitutiveLawEarthquake2019,dublanchet2022seismicity}, showing that the fundamental transition between stress-controlled and stress-rate-controlled modulation remains robust. Some studies incorporated fluid effects into unstable RSF systems (VW, $k<k_c$). \textcite{beeler2018constraints} proposed that under undrained conditions, dilatancy strengthening can stabilize the event patch and explain the weak tidal normal-stress modulation observed for LFEs on the San Andreas Fault. \textcite{zhao2025tidal} further showed that the introduction of  competition between hydraulic diffusion timescales and seismicity relaxation timescales can explain period-dependent tidal sensitivity.
Other studies directly resolved slip velocity response in unstable VW systems, including finite-fault simulations \parencite{ader2014response} and spring-block models \parencite{mercuri2025effects}. These studies showed that the response depends strongly on the ratio between the perturbation period and intrinsic RSF frictional evolution timescales scaling with $d_c/V_{\rm ss}$, where $d_c$ is the characteristic slip distance, $V_{\rm ss}$ is the background slip velocity.

Another line of work focused on faults undergoing stable sliding in the absence of perturbations. One class of models investigated how harmonic stress perturbations modulate stable sliding on VS faults. In this framework, harmonic stress perturbation primarily affects the slip velocity of the surrounding creeping zone, which subsequently indirectly triggers tremor and LFEs on embedded VW patches. \textcite{ader2012role} showed that the modulation response depends strongly on the characteristic RSF friction evolution timescale scaling with $d_c/V_{\rm ss}$. This interpretation was later applied to LFEs on the central San Andreas Fault by \textcite{beeler2013inferring}, and further extended to include pore pressure evolution and hydraulic diffusion by \textcite{sakamoto2022frictional}.

A second class of studies considers VW systems undergoing stable sliding ($k\gtrsim k_c$). \textcite{perfettini2001frictional} showed that harmonic stress perturbation can produce resonance-like amplification when the perturbation period approaches the characteristic resonance period scales as $T_c = 2\pi \sqrt{a/(b-a)}\, d_c/V_{\mathrm{ss}}$. This mechanism can potentially lead to direct triggering of slip events. This resonance behavior has been demonstrated experimentally by \textcite{boettcher2004effects,pignalberi2024effect} and invoked to interpret frequency-dependent sensitivity of fault slip and seismicity under tidal, seasonal, and hydrological perturbation \parencite{lowry2006resonant,panda2018seasonal,senapati2024gravity,sahoo2024tidal}. However, most previous studies primarily focused on explaining observed modulation phenomena using the resonance mechanism, whereas the physical conditions under which harmonic stress perturbations can directly trigger observable slip events remain poorly constrained. Building on the resonance mechanism identified by \textcite{perfettini2001frictional}, the present study systematically investigates the triggering behavior of VW faults subjected to harmonic stress perturbations using a spring-block RSF model. Rather than focusing only on the existence of resonance itself, we aim to identify the triggering regimes, characterize the correlations between perturbations and triggered events, and connect these behaviors to observations of tidal modulated slow earthquakes.

In this study, we first examine the transient response of stable sliding VW faults to instantaneous normal stress perturbations in order to characterize their fundamental response behavior (Section~\ref{sec:Fundamental response of VW stable faults}). We then investigate the response to harmonic stress perturbations representative of tidal perturbation (Section~\ref{sec:harmonic pertubations}). A nondimensional framework is developed to identify the key parameters controlling the coupling between external stress perturbations and intrinsic fault frictional properties (Section~3.2). Using spring-block simulations, we systematically explore slip velocity responses across this parameter space and identify the controlling parameter regimes under which harmonic stress perturbations can trigger slip events (Section~3.3). Then, we discuss the relationship between triggered events and tidal stress perturbations in the model, assess the robustness of these results to additional model parameters, and examine our model to tidal correlations observed in natural slow earthquakes in section~\ref{sec:discussion}. Finally, the main findings and their implications are summarized in section~\ref{sec: Conclusions}.

\section{Fundamental response of stable sliding VW faults to transient perturbations}
\label{sec:Fundamental response of VW stable faults}
A harmonic perturbation can be approximately represented by an idealized boxcar perturbation, which can further be decomposed into two consecutive step-like perturbations of opposite sign. Motivated by this simplified representation, we first investigate the response of a stable sliding VW fault to \rev{a} step-like and \rev{a} boxcar of finite duration normal stress perturbations to understand the fundamental response of a stable VW fault.

\subsection{Governing equations of the spring-block model}
\label{subsec:Governing equations}
\begin{figure}
    \centering
    \includegraphics[width=0.5\linewidth]{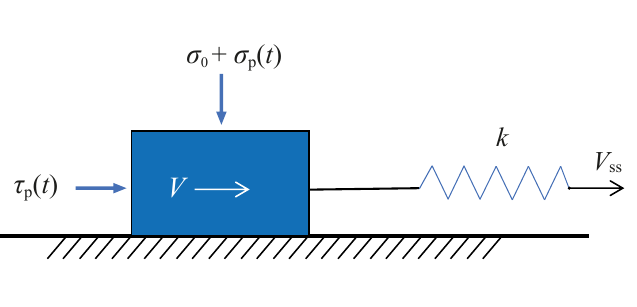}
    \caption{Spring–block model under stress perturbations. The block slides with velocity $V$ under a constant normal stress $\sigma_0$, subject to an imposed normal stress perturbation $\sigma_{\mathrm{p}}(t)$ and an applied shear stress perturbation $\tau_{\mathrm{p}}(t)$. The elastic loading is provided by a spring of stiffness $k$, driven at a constant velocity $V_{\mathrm{ss}}$.}
    \label{fig:spring_block_model}
\end{figure}

We employ a spring–block model under stress perturbations (Figure \ref{fig:spring_block_model}). Under a constant normal stress $\sigma_0$, the fault, modeled as a point and elastically loaded by a spring, is subjected to normal stress perturbations $\sigma_{\mathrm{p}}(t)$ (positive in compression) and shear stress perturbations $\tau_{\mathrm{p}}(t)$ (positive in the direction of slip), resulting in a slip velocity $V$. We restrict our analysis to a quasi-dynamic spring–slider model, as our primary goal is to identify the conditions under which harmonic perturbations can trigger slip events \rev{and to characterize the resulting slip velocity response}. 

The linear momentum balance of the spring-block system gives
\begin{align}
    \tau  &= \tau_0 + k \,V_{\mathrm{ss}}\, t  - k \, \delta - \eta V + \tau_p(t), 
    \label{eq:eq_sb} \\
    \sigma  &= \sigma_0 + \sigma_{\mathrm{p}}(t),  \label{eq:sigma1}
\end{align}
where $\tau$ is the resolved shear stress, $\tau_0$ is the initial shear stress at $t=0$ ($\delta = 0, V = 0$) and positive in the direction of slip, and $\tau_p(t = 0)  = 0$.  $\sigma$ is the resultant normal stress assuming compression is positive, $k$ is the spring stiffness, $V_{\mathrm{ss}}$ is the slip velocity of the loading point, $\delta$ is the slip of the block and $V={d\delta}/{dt}$ is the slip velocity of block. $\eta$ represents the radiation damping coefficient, which approximates inertial effects and prevents the slip velocity from becoming unbounded during instabilities, and is given by $\eta={\mu}/{2c_s}$, where $\mu$ is the shear modulus and $c_s$ is the shear wave velocity \parencite{rice1993spatio}.

We use the classical Rate-and-state dependent friction law (RSF) given by ~\textcite{dieterich1979a,ruina1983},
\begin{align}
    f(V, \theta)&= f_0 + a \log \left( \dfrac{V}{V_0} \right) + b \log \left( \dfrac{V_0 \theta}{d_c} \right), 
\end{align}
where, $f_0$ is a reference sliding friction coefficient when the fault is sliding at reference velocity $V_0$ determined from laboratory rock friction experiments~\parencite{marone1998laboratory}. $\theta$ is a state variable that characterizes the state of the sliding surfaces and has units of time. $d_c$ is the characteristic slip distance for the evolution of  $\theta$. $a$ represents the "direct effect", such that for fixed $\theta$, the friction increases logarithmically with slip velocity. $b$ characterizes the "evolution effect", describing the time-dependent weakening of friction through the evolution of the state variable. Under the RSF framework, we assume that the resolved shear stress is always equal to the frictional resistance and $\tau = f(V, \theta) \, \sigma$.

\begin{revblock}
Rate-and-state friction law (RSF) commonly employ two widely used state evolution laws: the aging law \parencite{dieterich1979a} and the slip law \parencite{ruina1983}. Both formulations were originally developed to reproduce laboratory friction experiments, but they differ in their response to abrupt velocity steps \parencite[Figure~11.8]{segallEarthquakeVolcanoDeformation2010}. 
For the slip law, frictional weakening following a velocity jump occurs over approximately the same characteristic slip distance regardless of perturbation amplitude. For this reason, sufficiently large perturbations increase the weakening rates and thus  may destabilize the fault even if $k > k_c$, making it only conditionally stable~\parencite{Gu1984}. In contrast, under the aging law, the characteristic slip distance increases with increasing velocity jump amplitude but the weakening rate $\sim b\sigma/d_c$ remains the same, due to which VW fault remain unconditionally stable for $k > k_c$~\parencite{ranjith1999, ciardoNonlinearStabilityAnalysis2025}. In the present study, we focus on relatively small harmonic perturbations acting on regions where slow earthquakes are observed. Under these conditions, the aging and slip laws are expected to exhibit qualitatively similar responses.      
\end{revblock}

We therefore primarily present results using the aging law ~\parencite{dieterich1979a}
\begin{align}
    \dfrac{d\theta}{dt} &= 1 - \dfrac{V \theta}{d_c}, 
\end{align}
in the subsequent sections, \rev{while the corresponding slip law results are discussed further in Section \ref{sec:additional_papameters}.}

In this study, we focus on a stable sliding VW fault, so we adopt $a/b = 0.9$ with $k/k_c=1.1$. Although both normal stress  $\sigma_{\mathrm{p}}(t)$ and shear stress $\tau_{\mathrm{p}}(t)$ perturbations are considered in the nondimensionalization (Section~\ref{subsec:non_demensionalization}), all subsequent simulations involving step, box and harmonic normal stress perturbations set $\tau_{\mathrm{p}}(t)=0$ for simplicity. For the step-like and boxcar perturbations in Sections~\ref{subsec:step_change} and~\ref{subsec:box_change}. \rev{The perturbation is applied to a fault in steady state sliding, with $V=V_{\mathrm{ss}}$ and $\theta=d_c/V_{\mathrm{ss}}$ as the initial condition.} Other model parameters are summarized in Table~\ref{tab:parameters}.

\begin{revblock}
\subsection{Resonance-like slip velocity amplification in RSF spring-block models}

 \textcite{perfettini2001frictional} theoretically described  a resonance phenomenon in a velocity weakening fault undergoing stable sliding due to small harmonic perturbations in either normal or shear stress. Due to this resonance, a fault system whose spring stiffness is just above its critical stiffness can produce a strong amplification of the slip velocity.

Assuming small harmonic perturbations about stable steady sliding with $k>k_c$, \textcite{perfettini2001frictional} linearized the RSF spring-block equations with harmonic perturbations around the steady state, and derived the following relation for change in the slip velocity $\Delta V$ \parencite[Eq.~B4]{perfettini2001frictional}
\begin{equation} \label{eq:perfreson}
\Delta V
=
q\,V_{\rm ss}
\frac{
q[\Delta\sigma(\mu_{ss}-\alpha)-\Delta\tau]
-i(\Delta\sigma\mu_{ss}-\Delta\tau)
}{
kd_c-a\sigma_0q^2+iqd_c(k-k_c)
},
\end{equation}
where $q=2\pi d_c/(TV_{\rm ss})$. The resonance-like amplification of the slip velocity occurs when denominator above becomes extremely small. The imaginary part, $iqd_c(k-k_c)$, measures the distance from critical stiffness and therefore decreases as $k$ approaches $k_c$. The remaining real part vanishes when $k_c d_c-a\sigma_0 q^2$ goes to zero which determines the characteristic resonance period given by
$
T_c
=
{2\pi d_c}/{V_{\rm ss}}
\left(\sqrt{{a}/{(b-a)}}\right)$. This time scale is same as the period of quasi-static, self-sustained oscillations of slip velocity that arise when $k \sim k_c$~\parencite[Eq.~24,32]{rice1985} in the unperturbed spring-block model.

It is important to emphasize that, while the Linker–Dieterich effect does modify the response to normal stress perturbations (as discussed in \ref{sec:additional_papameters}), the resonance phenomenon itself does not intrinsically depend on either the Linker–Dieterich parameter or even normal stress perturbations. As is evident from numerator of  Eq.~\eqref{eq:perfreson}, resonance behavior also arises for purely harmonic shear stress perturbations with no normal stress change ($\Delta\sigma=0$). Thus, the slip velocity amplification is controlled primarily by the denominator $kd_c-a\sigma_0q^2+iqd_c(k-k_c)$, which decides the essential ingredients for the resonance (i) VW fault  with $k\gtrsim k_c$ and (ii) a correspondence in timescales between the perturbation period $T$ and the characteristic resonance period $T_c$.
\end{revblock}

\subsection{Response to a step change in normal stress} 
\label{subsec:step_change}
We examine the response of a stable sliding VW fault to transient step changes in normal stress, prescribed as
$\sigma_{\mathrm{p}}(t)=\Delta\sigma H(t)$,
where $H(t)$ is the Heaviside function. Both downward (tensile, $\Delta\sigma<0$) and upward (compressive, $\Delta\sigma>0$) step changes are considered. As the spring stiffness exceeds the critical stiffness, for both types of step changes the system exhibits a common response pattern: following the stress perturbation, the slip velocity displays a damped oscillatory evolution with progressively decaying amplitude toward stable sliding, as illustrated by the inset in Fig.~\ref{fig:step}a. Similar damped oscillatory behavior is also observed for a boxcar perturbation. Here, we focus in the following on the maximum slip velocity induced by each stress perturbation, which for step changes corresponds to the first peak in slip velocity following the perturbation.

\subsubsection{Response to a tensile step change}
\begin{figure*}
    \centering
    \includegraphics[width=\textwidth]{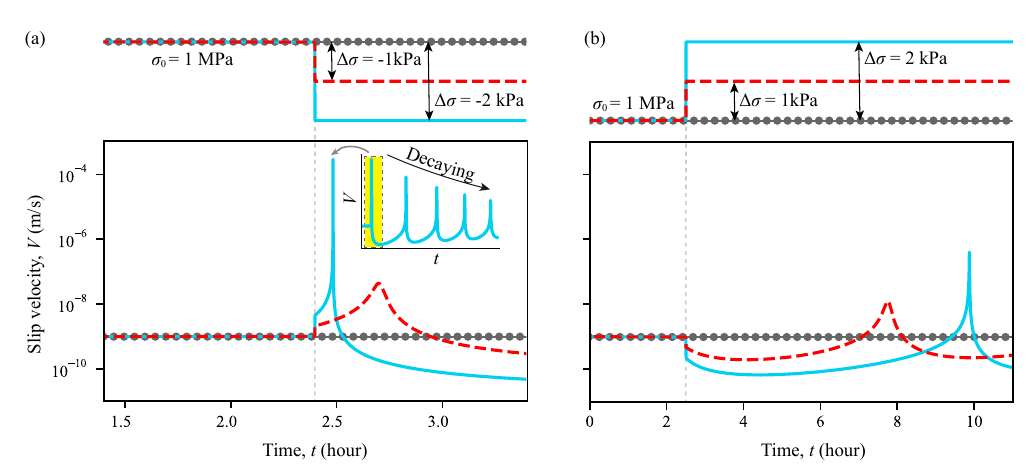}
    \caption{Responses of slip velocity to step changes in normal stress. (a) Downward step change and (b) Upward step change. Red dashed and blue solid curves correspond to $|\Delta\sigma| = 1~\mathrm{kPa}$ 
and $2~\mathrm{kPa}$, respectively. The dark gray line with circles indicates the reference case without stress perturbation. \rev{The inset in panel (a) shows the long-term damped oscillatory evolution of slip velocity $V$ as a function of time $t$.} \rev{It illustrates the gradual return toward stable sliding following the transient step perturbation.} The other model parameters are listed in Table~\ref{tab:parameters}.}
    \label{fig:step}
\end{figure*}

Figure~\ref{fig:step}a shows the slip velocity response to a downward step change in normal stress ($\Delta\sigma<0$). The response consists of two stages: an instantaneous increase in slip velocity at the moment of stress change (see \ref{app:Instantaneous response of fault to step change} for theoretical analysis of this instantaneous response), followed by a gradual acceleration toward a peak velocity. As the perturbation amplitude increases, the peak velocity becomes larger and is reached over a shorter time. This behavior is consistent with the results of \textcite{paul2024frictional}, who studied failure of creeping landslide sliding on the RSF frictional interface due to periodic pressure perturbations using a block model without an elastic spring. In contrast, our model includes an elastic spring under the stable condition ($k > k_c$), which allows the slip velocity to recover after the peak, thereby preventing runaway instabilities.

\begin{table*}[t]
\centering
\caption{Parameters used in numerical simulations of step and boxcar normal stress perturbation.}
\label{tab:parameters}
\begin{tabular}{lccc}
\hline
\textbf{Parameter} & \textbf{Symbol} & \textbf{Value} & \textbf{Unit} \\
\hline
Reference stable sliding friction coefficient & $f_0$ & $0.6$ & \\
Reference velocity & $V_0$ & $10^{-6}$ & $\rm m/s$ \\
RSF parameters & $a/b$ & $0.9$ & \\
Direct effect coefficient & $a$ & $0.0008$ & \\
Loading velocity & $V_{\mathrm{ss}}$ & $10^{-9}$ & $\rm m/s$ \\
Characteristic slip distance & $d_{c}$ & $1\times 10^{-6}$ & $\rm {m}$\\
Background (constant) normal stress & $\sigma_{0}$ & $1$ & $\rm {MPa}$\\
Radiation damping coefficient & $\eta$ & $8\times10^{5}$ & $\rm Pa \cdot s/m$ \\
Spring stiffness & $k$ & $9.8\times10^{7}$ & ${\rm Pa/m}$ \\
\hline
\end{tabular}
\end{table*}

\subsubsection{Response to a compressive step} 
\label{Response to an upward step}
Figure~\ref{fig:step}b shows the response to an upward step change in normal stress ($\Delta\sigma>0$). In this case, the slip velocity decreases instantaneously at the time of the stress change, followed by an evolution phase during which the velocity reaches a local minimum before accelerating toward a peak. Larger compressive perturbations lead to higher peak velocities that occur after longer times.

The contrasting (\rev{asymmetric}) responses to tensile and compressive step changes reflect the opposite instantaneous effects of normal stress perturbations on frictional resistance. A sudden increase in normal stress strengthens the fault while the shear stress remains unchanged, causing the system to become more locked and delaying acceleration toward the peak slip velocity. In contrast, a tensile step reduces frictional resistance, facilitating slip and shortening the time required to reach the peak slip velocity.

\subsection{Response to a boxcar change in normal stress} 
\label{subsec:box_change}
In this section, we examine the response of a stable sliding VW fault to finite-duration normal stress perturbations, prescribed as boxcar perturbations. A boxcar perturbation consists of two consecutive step changes of opposite polarity separated by a finite duration, $\sigma_{\mathrm{p}}(t) = \Delta \sigma \left[ H(t - t_1) - H(t - t_2) \right]$,
where $H(t)$ is the Heaviside function, $t_1 < t_2$, and $T_{\mathrm{box}} = t_2 - t_1$ denotes the duration of the perturbation. Depending on the sign of the initial step, the perturbation corresponds to either a box-up ($\Delta\sigma > 0$) or a box-down ($\Delta\sigma < 0$) loading. Here, we fix $|\Delta\sigma| = 1~\mathrm{kPa}$. As in the step-change case (Section~\ref{subsec:step_change}) and we characterize the fault response by the first peak slip velocity following the termination of the boxcar perturbation.

\subsubsection{Response to a compressive change} 
\begin{figure*}
    \centering
    \includegraphics[width=\textwidth]{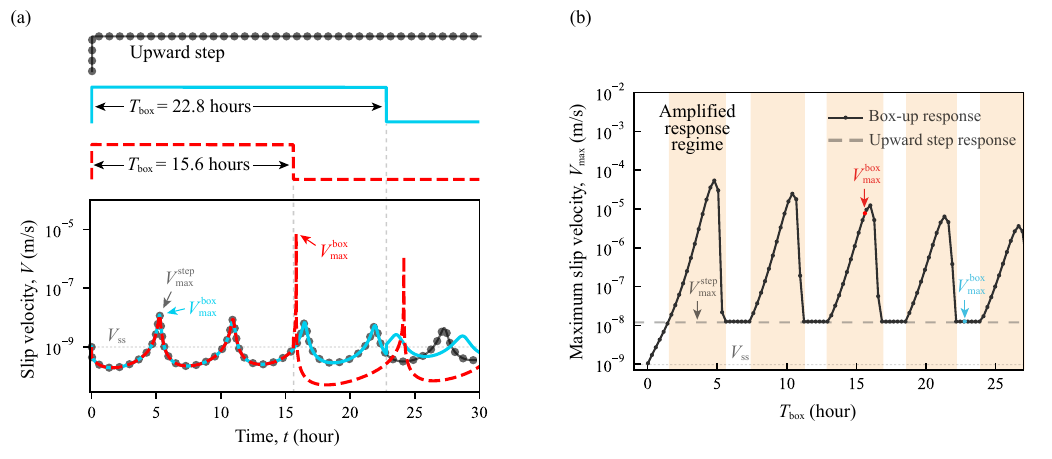}
    \caption{
    Responses of slip velocity to a box-up change in normal stress with $|\Delta\sigma| = 1.0~\mathrm{kPa}$ (the other model parameters are listed in  Table~\ref{tab:parameters}).
    (a) Upper: imposed normal stress histories for $T_{\mathrm{box}}=22.8~\mathrm{hours}$ (blue) and $15.6~\mathrm{hours}$ (red), compared with a single upward step (dark gray line with circles).
    Lower: corresponding slip velocity responses.
    (b) Maximum slip velocity $V_{\max}$ as a function of $T_{\mathrm{box}}$. The shaded region highlights the range of $T_{\mathrm{box}}$ associated with amplified responses compared to the upward step change.}
    \label{fig:box_up}
\end{figure*}

We first consider the response to a box-up perturbation ($\Delta\sigma > 0$), in which an upward (compressive) step in normal stress is followed by a downward (tensile) step after a finite duration $T_{\mathrm{box}}$. Figure~\ref{fig:box_up}a shows the imposed normal stress histories for $T_{\mathrm{box}} = 15.6~\mathrm{hours}$ and $22.8~\mathrm{hours}$, together with the reference response to an upward step. The corresponding slip velocity responses are shown in the lower panel of Figure~\ref{fig:box_up}a. 

As indicated by the dark gray curve with circles, an upward step change produces an initial peak in slip velocity, followed by a gradual decay toward stable sliding. This first peak defines the maximum slip velocity associated with the step response, $V_{\max}^{\mathrm{step}} \sim 10^{-8}~\mathrm{m/s}$. For $T_{\mathrm{box}} = 15.6~\mathrm{hours}$ (red curve), the first peak slip velocity following the termination of the box-up perturbation reaching values of the order of $10^{-5}~\mathrm{m/s}$, significantly larger than $V_{\max}^{\mathrm{step}}$. After this peak, the slip velocity again decays gradually, so that this post-termination peak defines the maximum slip velocity for this box-up case, $V_{\max}^{\mathrm{box}}$. In contrast, for $T_{\mathrm{box}} = 22.8~\mathrm{hours}$ (blue curve), the first peak slip velocity following the termination of the box-up perturbation is lower than $V_{\max}^{\mathrm{step}}$, and the slip velocity subsequently decays. As a result, the maximum slip velocity in this case is given by the step-response peak, $V_{\max}^{\mathrm{step}}$.

To generalize these observations, Figure~\ref{fig:box_up}b shows the maximum slip velocity $V_{\max}$ as a function of $T_{\mathrm{box}}$. 
For a fixed perturbation amplitude, an increase in slip velocity, relative to the step change response, occurs only over a limited range of perturbation periods, highlighted by the shaded region in the figure. Outside this range, the maximum velocity remains comparable to or smaller than $V_{\max}^{\mathrm{step}}$.

\begin{figure*}
    \centering
    \includegraphics[width=\textwidth]{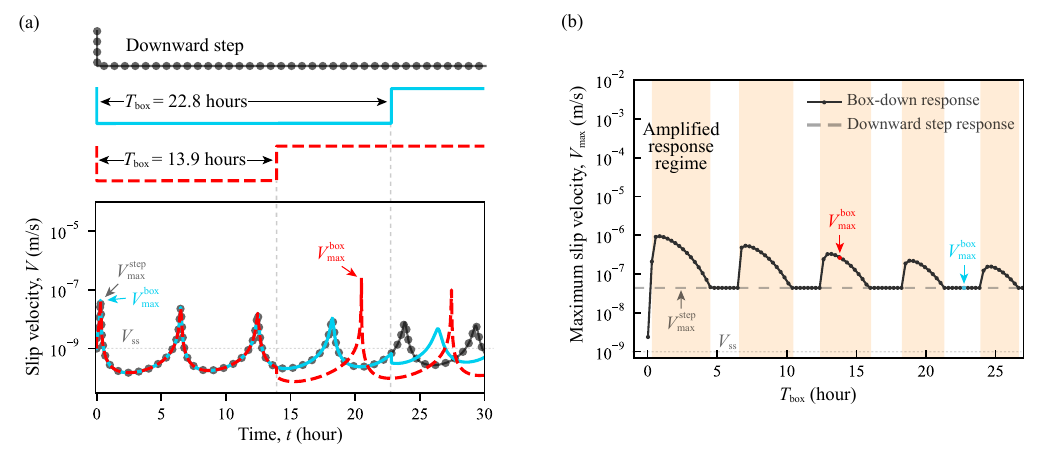}
    \caption{
    Responses of slip velocity to a box-down change in normal stress with $|\Delta\sigma| = 1.0~\mathrm{kPa}$.
    (a) Upper: imposed normal stress histories for $T_{\mathrm{box}}=22.8~\mathrm{hours}$ (blue) and $13.9~\mathrm{hours}$ (red), compared with a single downward step (dark gray line with circles).
    Lower: corresponding slip velocity responses.
    (b) Maximum slip velocity $V_{\max}$ as a function of $T_{\mathrm{box}}$. }
    \label{fig:box_down}
\end{figure*}

\subsubsection{Response to a tensile change} 
We next consider the response to a  box-down perturbation ($\Delta\sigma < 0$), in which a downward (tensile) step in normal stress is followed by an upward (compressive) step after a finite duration. The imposed normal stress histories for $T_{\mathrm{box}} = 13.9~\mathrm{hours}$ and $22.8~\mathrm{hours}$ are shown in the upper panel of Figure~\ref{fig:box_down}a, together with the reference response to a downward step. The corresponding slip velocity evolutions are shown in the lower panel.

As indicated by the dark gray curve with circles, a downward step produces a peak in slip velocity followed by a gradual decay toward stable sliding, defining the maximum slip velocity associated with the step response, $V_{\max}^{\mathrm{step}}$. For $T_{\mathrm{box}} = 13.9~\mathrm{hours}$ (red curve), the post-termination peak exceeds the step response peak, whereas for $T_{\mathrm{box}} = 22.8~\mathrm{hours}$ (blue curve), the post-termination peak remains lower than $V_{\max}^{\mathrm{step}}$. Figure~\ref{fig:box_down}b summarizes the maximum slip velocity $V_{\max}$ as a function of $T_{\mathrm{box}}$. 
As in the box-up case, the amplification of slip velocity relative to the step response occurs only for a limited range of perturbation periods, although the degree of amplification and the corresponding $T_{\mathrm{box}}$ interval differ. Outside this range, the maximum slip velocity is controlled by the step response.

\subsubsection{Summary: Resonance and slip velocity amplification}
The analysis of step and finite-duration (boxcar) normal stress perturbations illustrates how transient stress perturbations effect the slip velocity response on a stable sliding VW fault. Importantly, for a stable sliding VW fault, the slip velocity response is strongly dependent on both the perturbation amplitude $\Delta\sigma$ and its duration $T_{\mathrm{box}}$. Within some finite range of $T_{\mathrm{box}}$, boxcar perturbations lead to a pronounced amplification of slip velocity relative to the corresponding step response, with peak velocities reaching up to several orders larger than those of the corresponding response of the step change. Outside this range, the maximum slip velocity is largely controlled by the step-like response and no significant amplification occurs. This selective amplification indicates that the slip response is strongly enhanced only for specific perturbation durations. Such behavior is consistent with a resonance-like response and motivates the examination of harmonic stress perturbations in the following section. In the above section, results for single step and boxcar inputs have been shown for normal stress perturbations as an illustration. However, the same analysis can also be carried out for shear stress perturbations, and the fundamental response will remain the same.

\section{Harmonic stress perturbation response of stable sliding VW faults}
\label{sec:harmonic pertubations}

Building on the resonance-like behavior identified for step and boxcar perturbations, we now extend our analysis to harmonic stress perturbations. Our goal is to determine the conditions under which such harmonic perturbation can generate observable slip events on a stable sliding VW fault. Of particular note, in stress perturbation problems, the system response does not simply scale with the perturbation amplitude or period, but exhibits marked sensitivity to the fault friction parameters. To capture this dependence, we first carry out a nondimensional analysis to identify the key controlling parameters that encapsulate the interplay between fault properties and perturbation characteristics. We then use numerical simulations to systematically explore the slip velocity response across this parameter space, thereby identifying the parameter regimes in which harmonic stress perturbations can trigger slip events whose slip velocities exceed a prescribed threshold and are therefore potentially observable.

\subsection{Scaling and nondimensional parameters}
\label{subsec:non_demensionalization}
To identify the key factors controlling tidal triggering, we consider harmonic, in-phase perturbations in both normal and shear stresses, expressed as
$$
\sigma_{\rm p}(t) = \Delta\sigma\sin(2\pi t/T),\qquad
\tau_{\rm p}(t) = \Delta\tau\sin(2\pi t/T),
$$
where $T$ is the tidal stress perturbation period, $\Delta \sigma > 0$  is the amplitude of tidal normal stress, and $\Delta \tau > 0$ is the amplitude of tidal shear stress.

Building on the above model, we perform a nondimensional analysis which is explained in detail in~\ref{app:nondimensional analysics}. The resulting formulation yields six nondimensional parameters ($R_{ab}$, $\kappa$,  $\mathcal{N}$, $\epsilon$, $P_{\sigma}$ and $P_T$), each playing a critical role in governing the slip behavior. These nondimensional parameters characterize the long-term response to persistent harmonic perturbations, where the influence of the initial conditions becomes negligible.

The first nondimensional parameter is 
\begin{align}
    R_{ab} = \dfrac{a}{b} 
    \label{eq:R_{ab}}
\end{align}

which characterizes the relative contribution of the "direct effect" ($a$) and the "evolution effect" ($b$) of the friction in the RSF framework. The frictional response depends on the ratio $R_{ab}$, which governs whether steady state friction increases or decreases with the slip velocity. VS behavior occurs when $R_{ab} > 1$, velocity neutral behavior when $R_{ab} \rightarrow 1$, and VW behavior when $0 < R_{ab} < 1$. Laboratory experiments indicate that $R_{ab}$ typically exceeds $0.9$ for VW surfaces \parencite{kilgore1993velocity,blanpied1998quantitative}. Moreover, in subduction zones, where constitutive properties vary primarily with temperature, the value of $R_{ab}$ is expected to be close to $1$ in the transition zone between locked and creeping regions, that is, between VW and VS regimes. 
This parameter $R_{ab}$ is directly related to the quantity $R_b = (b-a)/b$ introduced by \textcite{barbot2019slow}.
This parameter can also be heuristically related to the ratio between the stress drop, $\tau_{0}-\tau_{r} \sim (b-a)\,\sigma_{0}$,
and the frictional strength drop, $\tau_{p}-\tau_{r} \sim b\,\sigma_{0}$~\parencite{madariaganotes1998, ericksonModelAperiodicityEarthquakes2008}, and commonly used S-ratio associated with the slip weakening friction law, $\mathcal{S} = (\tau_{p}-\tau_{0})/(\tau_{0}-\tau_{r}) \sim a/(b-a) = R_{ab}/(1 - R_{ab})$ \parencite{das1976, andrewsRuptureVelocityPlane1976}.

The second nondimensional parameter is 
\begin{align}
    \kappa = \dfrac{k}{k_c}
    \label{eq:kappa}
\end{align}

which controls the slip stability. \textcite{ruina1983} showed that the critical spring stiffness is $k_c = {(b - a) \, \sigma_0}/{d_c}$. The system undergoes stable slip when $\kappa>1$  and is unstable when $\kappa<1$, showing stick-slip behavior . 

The third nondimensional parameter is 
\begin{align}
     \mathcal{N} = \dfrac{ V_{\rm dyn}}{V_{\rm ss}}
    \label{eq:N}
\end{align}

which measures how far the stable sliding velocity is from the dynamic slip velocity scale at which radiation damping becomes important. Equivalently, $\mathcal{N}$ gives the approximate velocity amplification, $V/V_{\rm ss}$, required for radiation damping to become significant. Here $V_{\rm dyn}=a\sigma_0/\eta$ is the slip velocity scale at which the radiation damping stress $\eta V$ becomes comparable to the direct effect frictional strength scale $a\sigma_0$ \parencite[Eq.~A.8]{rubin2005earthquake}. Thus, large $\mathcal{N}$ indicates that radiation damping is negligible near stable sliding and becomes important only when $V/V_{\rm ss}\sim\mathcal{N}$.

The fourth nondimensional parameter is 
\begin{align}
      \epsilon = \dfrac{\Delta \sigma}{\sigma_0}
      \label{eq:epsilon}
\end{align}
which represents the ratio of tidal normal stress to the background normal stress \parencite{perfettini2001frictional}. Since the tidal normal stress is of the order of kPa, and the background normal stress is usually of the order of MPa, this parameter is much smaller than 1, and it is not the controlling parameter for tide-induced triggering. 

The fifth nondimensional parameter is
\begin{align}
     P_{\sigma} = \dfrac{|\Delta \tau\, -f_*^{\rm ss} \, \Delta \sigma\,|}{a \sigma_0}
    \label{eq:P_sigma}
\end{align}
which represents a normalized perturbation amplitude. The term $f_*^{\rm ss}$ denotes the steady state friction coefficient at the background slip velocity $V_{\rm ss}$, is given by $f_*^{\rm ss} = f_0 + (a-b)\log\,\left({V_{\mathrm{ss}}}/{V_0}\right)$. $|\Delta \tau\, -f_*^{\rm ss} \, \Delta \sigma\,|$ is the magnitude of tidal Coulomb stress perturbation in the direction of slip in the context of the RSF.  The tidal stress is a persistent perturbation and alternates periodically between positive and negative values. We are only interested in the long-term slip velocity response of the fault and the sign of the tidal Coulomb stress perturbation is inconsequential. The $a\sigma_0$ is the instantaneous frictional resistance offered by the fault. Thus, $P_\sigma$ represents the ratio of Coulomb stress change and the instantaneous frictional strength change due to tidal perturbation. $P_\sigma$ is closely related to the parameter
$\mathcal{T} = (f_s \sigma_{0} - \tau_{0}) /(f_s \Delta p_c)$
used in studies of induced seismicity, where
$P_{\sigma} \sim 1/\mathcal{T}$
\parencite{
garagashNucleationArrestDynamic2012,saezThreedimensionalFluiddrivenStable2022}.
The parameter $\mathcal{T}$ represents the ratio between the distance to failure $ (f_s \sigma_{0} - \tau_{0})$ and the Coulomb stress change induced by fluid pressure $f_s \Delta p_c$.
Under this interpretation, the quantity $a \sigma_{0}$ in RSF
plays a role analogous to the distance to failure ($f_s \sigma_{0} - \tau_{0})$,
where $f_s$ is the static friction coefficient, $\tau_{0}$ and $\sigma_{0}$ are the in-situ shear stress and background (or constant) normal stress,
respectively. A smaller value of $a \sigma_{0}$ therefore implies that the fault is more susceptible to small stress perturbations in RSF framework.

The last nondimensional parameter is
\begin{align}
P_T = \dfrac{T}{t_*} = \dfrac{T\,V_{\mathrm{ss}}}{d_c}
\label{eq:P_T}
\end{align}
which represents the perturbation period normalized by the characteristic friction evolution timescale $t_* = d_c / V_{\mathrm{ss}}$.
Here, $T$ denotes the period of the imposed harmonic stress perturbation, and $t_*$ characterizes the timescale over which the state variable evolves in response to changes in slip velocity \parencite{perfettini2001frictional,ader2012role,paul2024frictional}. In VW systems ($k \gtrsim k_c$), the characteristic resonance period derived by \textcite{perfettini2001frictional},
$T_c = 2\pi \sqrt{a/(b-a)}\, d_c/V_{\mathrm{ss}}$,
scales directly with this characteristic friction evolution timescale, which motivates the use of $P_T$ as a key parameter for describing resonance-like triggering in the present study.
When $P_T \ll 1$, the perturbation period is short compared to the friction evolution timescale, so the state variable remains effectively frozen over a perturbation cycle. In this regime, the fault response is dominated by the direct velocity effect, characterized by the parameter $a$. In contrast, for $P_T \gg 1$, the perturbation varies slowly relative to state evolution, allowing the state variable to evolve during each cycle. 

Analogously to $P_T = T/t_*$, the perturbation period can also be normalized by the characteristic seismicity response timescale as
\begin{equation}
P_{Ta} = \dfrac{T}{t_a}=\dfrac{T k V_{\mathrm{ss}}}{a \sigma_0},
\end{equation}
where $t_a = a \sigma_0/(k V_{\mathrm{ss}})$ corresponds to the Dieterich characteristic seismicity relaxation timescale when the background stressing rate is written as $kV_{\mathrm{ss}}$ \parencite{dieterich1994constitutive,dieterich2007applications}. This timescale controls the temporal evolution of seismicity rates under stress perturbations and has been widely used to describe stress-controlled and stress-rate-controlled modulation under harmonic perturbation \parencite{beeler2003earthquakes,heimisson2018constitutive,heimisson2020analytical}.

Among these parameters, the first three ($R_{ab}$, $\kappa$, $\mathcal{N}$) play key roles in controlling fault slip behavior and are already well understood in the absence of any external perturbations \parencite{barbot2019slow,wangQuantitativeCharacterizationSimulated2024}. The remaining two parameters, $P_{\sigma}$ and $P_T$, therefore emerge as the primary control parameters governing the fault response to harmonic stress perturbations. For fixed $\kappa=k/k_c$ and $R_{ab}=a/b$, $P_{Ta}=P_T\,\kappa(1-R_{ab})/R_{ab}$. Although $\epsilon = \Delta\sigma / \sigma_0$ appears explicitly in the governing equations, we find that the slip velocity response is only weakly sensitive to $\epsilon$ within the small perturbation regime ($\epsilon \ll 1$) considered here. Accordingly, $\epsilon$ is fixed in the following analysis, allowing us to focus on the dominant control parameters $P_{\sigma}$ and $P_T$.

\begin{table}[htbp]
\centering
\caption{Model parameters used in numerical simulations of harmonic normal stress perturbations}
\label{tab:non_dimensional_parameters}
\begin{tabular}{lccc}
\hline
\textbf{nondimensional Parameter} & \textbf{Symbol} & \textbf{Value}\\
\hline
Friction parameter & $R_{ab}$ & $0.9$ \\
Stiffness & $\kappa$ & $1.1$ & \\
Radiation damping & $\mathcal{N}$ & $10^{6}$  \\
Stress perturbation & $\epsilon$ & $10^{-3}$  \\
Perturbation amplitude & $P_{\sigma}$ & $0.1-1$  \\
Perturbation period & $P_T$ & $1-100$  \\
\hline
\end{tabular}
\end{table}

\subsection{A phase diagram of slip behavior due to perturbations} 
\label{subsec:tidal_change_vmax}

We systematically explore the slip velocity response to harmonic stress perturbations over a broad range of $P_\sigma$ and $P_T$. Our simulations over a broader parameter range show that cases with $P_T<1$ and $P_T>100$ are generally dominated by small slip velocity oscillations that follow the stress perturbation. Based on these results, we focus the main parameter space exploration on the range $P_T \in [1,100]$, where amplified slip velocity responses are more likely to occur. For $P_\sigma>1$, kPa-scale tidal stresses would exceed $a\sigma_0$. This would require $a\sigma_0$ to be at most of kPa order, corresponding to either very low effective normal stress or unusually small values of the RSF "direct effect" parameter $a$. We therefore focus on the parameter space $P_{\sigma} \in [0.1,1]$ and $P_T \in [1,100]$. All other nondimensional parameters ($R_{ab}$, $\kappa$, $\mathcal{N}$, and $\epsilon$) are held constant (Table~\ref{tab:non_dimensional_parameters}). Because tidal stress amplitudes and periods are relatively well constrained (we adopt representative tidal values $\Delta\sigma = 1~\mathrm{kPa}$ and $T = 12~\mathrm{hours}$ here), varying $P_{\sigma}$ and $P_T$ effectively corresponds to exploring faults with different friction parameters.
Slip behavior is analyzed within the time window $t/t_* = 10000-20000$ to ensure statistically steady conditions under persistent harmonic perturbation. The slip velocity response is classified relative to the dynamic velocity scale $V_{\mathrm{dyn}}=\mathcal{N}V_{\mathrm{ss}}$. Fast events are identified when $V/V_{\mathrm{ss}}>\mathcal{N}$ (i.e., $V>V_{\mathrm{dyn}}$), slow events are identified when $10^{-3}\mathcal{N} \leq V/V_{\mathrm{ss}} \leq \mathcal{N}$. Slip velocities that remain below $10^{-3}\mathcal{N}$ are classified here as creeping, although they may still exhibit small slip velocity oscillations. Numerical implementation details, including the solver configuration and error control strategy, are provided in ~\ref{sec:Numerical Methods}.

\begin{figure*}
    \centering
    \includegraphics[width=\textwidth]{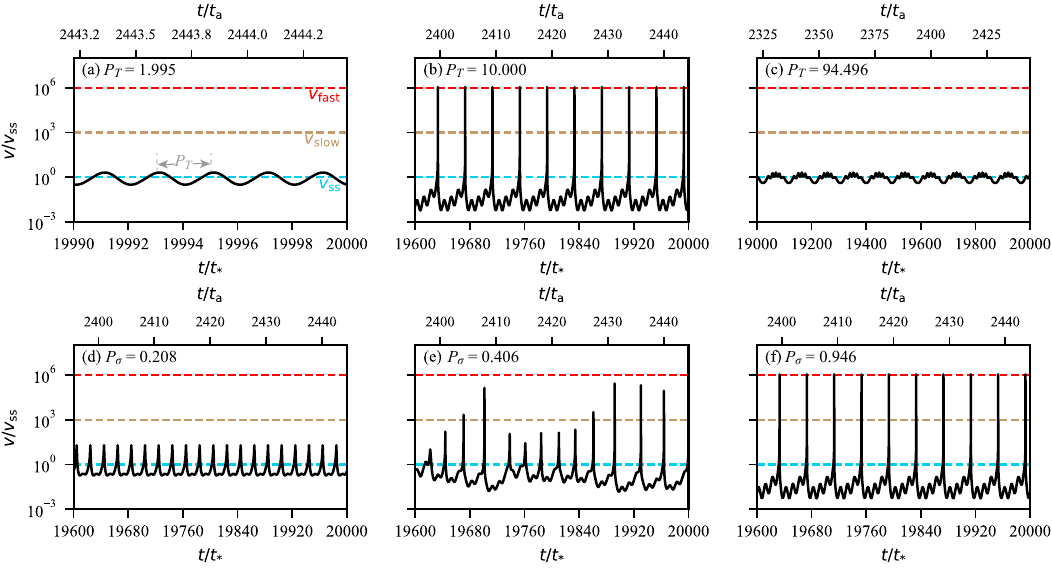}
   \caption{Normalized slip velocity $V/V_{\mathrm{ss}}$ as a function of time under harmonic stress perturbations. Panels (a)-(c) show cases with fixed perturbation amplitude $P_{\sigma}=0.892$ and increasing perturbation period $P_T=1.995$, $10.000$, and $94.496$, respectively, illustrating a transition from creeping to events and back to creeping. Panels (d)-(f) show cases with fixed perturbation period $P_T=10.000$ and increasing perturbation amplitude $P_{\sigma}=0.208$, $0.406$, and $0.946$, demonstrating progressively stronger slip velocity responses. Dashed horizontal lines indicate slip velocity thresholds. Time is normalized by $t_*$ (bottom axis) and $t_a$ (top axis).}
    \label{fig:PTPsigma_vary}
\end{figure*}

Figure~\ref{fig:PTPsigma_vary} illustrates representative normalized slip velocity responses under harmonic stress perturbations. Figures~\ref{fig:PTPsigma_vary}a–c show time series responses for different normalized perturbation periods $P_T$ at a fixed amplitude $P_{\sigma}=0.892$.
As $P_T$ increases, the slip velocity response evolves non-monotonically, transitioning from creeping to events and eventually returning to a creeping. 
For small $P_T$, the perturbation varies rapidly relative to the friction evolution timescale, limiting the response to small, nearly instantaneous velocity oscillations that closely track the imposed stress. At intermediate $P_T$, where the perturbation period becomes comparable to the friction evolution timescale, the response is strongly amplified, producing slip events indicative of resonance-like behavior. For large $P_T$, the perturbation varies slowly enough for the friction evolution timescale to fully adjust within each cycle, leading to quasi-static responses.

Figure~\ref{fig:PTPsigma_vary}d-f presents slip velocity responses for a fixed perturbation period $P_T=10$ and increasing perturbation amplitude $P_{\sigma}$. With increasing $P_{\sigma}$, the fault response evolves from creeping to slow events, and eventually to fast events, reflecting the progressively stronger influence of stress perturbations. In addition, intermediate amplitudes may produce complex or irregular slip velocity patterns (as illustrated by case e), suggesting that the system can exhibit nonlinear slip velocity response even under harmonic perturbation.

\begin{figure}
    \centering
    \includegraphics[width=\linewidth]{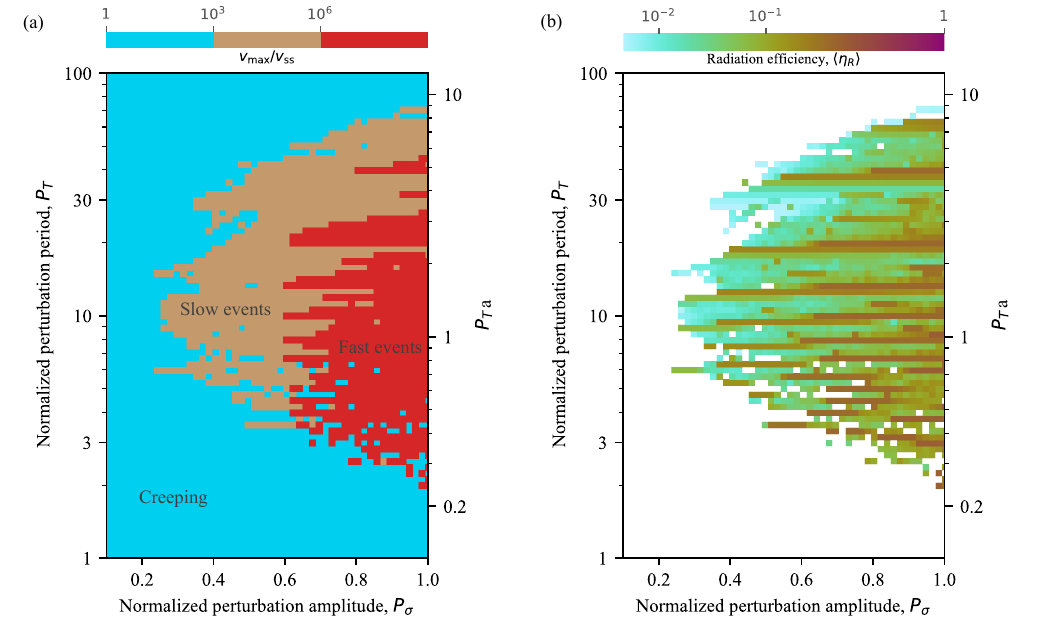}
    \caption{Response phase diagrams of a stable sliding VW fault in the $(P_{\sigma}, P_T)$ parameter space under harmonic normal stress perturbations. (a) Phase diagram colored by the normalized maximum slip velocity $V_{\max}/V_{\mathrm{ss}}$. Light blue indicates creeping behavior, brown denotes slow slip events, and red corresponds to fast events.(b) The same parameter space colored by the average radiation efficiency $\langle \eta_R \rangle$, computed only for cases where slip events occur. An event is defined as a slip episode with $V_{\max}/V_{\mathrm{ss}} > 10^{3}$, and $\langle \eta_R \rangle$ represents the mean value averaged over all events in each simulation.}
    \label{fig:phase_dagram_sbmodel}
\end{figure}

To synthesize the time series results described above, we summarize all the simulation results in the $(P_{\sigma}, P_T)$ phase diagram (Figure~\ref{fig:phase_dagram_sbmodel}a). The color scale represents the normalized maximum slip velocity, $V_{\max}/V_{\mathrm{ss}}$, which allows us to distinguish creeping (light blue), slow events (brown), and fast events (red). The phase diagram reveals the triggered events regimes, only within a confined region of parameter space, approximately $P_{\sigma} \gtrsim 0.2$ and $2 \lesssim P_T \lesssim 70$. Outside this window, the fault response remains creeping, with no detectable slip events. This result demonstrates that neither perturbation amplitude nor period alone is sufficient to trigger slip events; instead, both must fall within specific ranges. So the phase diagram provides a systematic and quantitative characterization of the conditions under which stress perturbations can trigger observable slip events on a stable sliding VW fault.

\subsection{The radiation efficiency of triggered events}
\label{subsec:tidal_change_radiation_efficiency}
 In the last section, we classified fast and slow events based on $V_{\mathrm{max}}/V_{\mathrm{ss}}$. Although the maximum slip velocity can indicate whether events occur under given stress perturbations, it does not provide insight into the detailed slip evolution, such as whether they are all fast or slow events. To address this limitation, we compute the average radiation efficiency, $\eta_R$, of all events. Here, we use the radiated energy ratio,
\begin{equation}
\eta_R = \frac{E_R}{\Delta W_o} \approx \frac{E_R}{D_o + E_R},
\end{equation}
where $E_R$ is the radiated seismic energy, $\Delta W_o$ is the elastic strain energy released during the event, and $D_o$ is the frictional energy dissipated during slip (see ~\ref{app:eta} for details). Thus, $\eta_R$ represents the fraction of released energy radiated as seismic energy \parencite{kostrovSeismicMomentEnergy1974} and provides an alternative description of slip velocity evolution. Fast events exhibit high $\eta_R$, as a large fraction of the released potential energy is converted into seismic waves, whereas slow events have low $\eta_R$ because almost all the available energy is dissipated on the fault. This distinction allows radiation efficiency also to provide insight into the relative proportion of fast versus slow events. Hereafter, we define events as slip events satisfying $V/V_{\mathrm{ss}} > 10^{-3}\mathcal{N}$ within the time window $t/t_* = 10000-20000$, and compute the average radiation efficiency over all such events in each catalog. Because $\eta_R$ spans several orders of magnitude, we use the geometric mean rather than the arithmetic mean: $\langle\eta_R\rangle = 10^{\langle\log \eta_R\rangle}$, where $\langle\log \eta_R\rangle$ is the average of $\log \eta_R$ over all events in each catalog.

Figure~\ref{fig:phase_dagram_sbmodel}b presents the phase diagram of the average radiation efficiency within the event area of Figure~\ref{fig:phase_dagram_sbmodel}a. We can find that larger $P_{\sigma}$ systematically yields higher radiation efficiency, indicating that a greater fraction of the released energy is radiated as seismic waves. The values of $P_T$ at which higher $\eta_R$ (purple) occurs exhibit a "comb-like" distribution. This behavior can be understood in terms of the response to box stress perturbations. As discussed in Section~\ref{subsec:box_change}, a box perturbation can be viewed as an upward stress step followed by a downward step. The timing of the downward step, which strongly influences the final response, is determined by the perturbation period $P_T$.
For very small $P_T$, the system is still in the rising phase of the upward step response and has not yet reached its first peak when the downward step occurs, so the combined perturbation remains too weak to exceed the event threshold. For very large $P_T$, the peak response to the upward step has already relaxed back toward stable sliding before the downward step is applied, leaving the combined perturbation again ineffective. Only for intermediate values of $P_T$ does the interaction between the upward and downward step responses produce sufficient amplification, giving rise to a series of discrete peaks in $\eta_R$ (Figures~\ref{fig:box_up}b and~\ref{fig:box_down}b).

\section{Discussion}
\label{sec:discussion}
\subsection{The correlation between triggered events and harmonic perturbations}
In our simulations, harmonic stress perturbations can trigger events on stable sliding VW fault due to resonance. whether the peak slip velocity of these events occurs preferentially near the tidal stress maximum, near the maximum tidal stressing rate, or shows no systematic phase relation. Addressing this issue is essential to assess whether such resonance-driven triggering could be identified in natural observations.

\begin{figure*}
    \centering
    \includegraphics[width=0.65\linewidth]{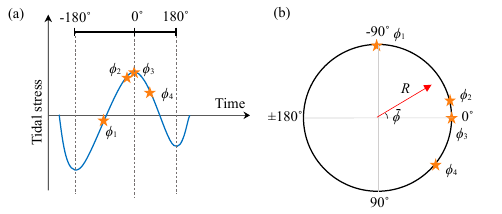}
    \caption{(a) Schematic showing the determination of the tidal phase $\phi$. In the calculation of tidal phase, the stress peak closest to the event origin time is defined as 0°, while the adjacent troughs are set to –180° and +180°. The intervals from –180° to 0° and from 0° to +180° are evenly divided in time, allowing each moment to be assigned a corresponding phase. (b) Determination of the mean tidal phase $\bar{\phi}$ and the resultant vector length $R$.}
    \label{fig:tidal_phase_define}
\end{figure*}

Similar to the approaches used to study tidal correlations in natural (slow) earthquakes \parencite{thomas2012tidal,royer2015tidal,van2016fortnightly,zhao2025tidal}, we quantify the correlation using the tidal phase distribution. The tidal phase $\phi$ indicates the position within the tidal stress cycle at which an event occurs, distinguishing whether the event takes place near a tidal stress peak, a trough, or during the rising or falling stage.
The definition of the tidal phase is illustrated in Figure \ref{fig:tidal_phase_define}a.

Because phase is a circular variable, the arithmetic mean is inappropriate (e.g., phase at $179^{\circ}$ and $-179^{\circ}$ are close, yet their arithmetic mean would misleadingly give $0^{\circ}$). So to characterize the phase distribution, we employ two standard parameters from circular statistics: the mean tidal phase $\bar{\phi}$ and the resultant vector length $R$. As shown in Figure~\ref{fig:tidal_phase_define}b, each phase $\phi$ is represented as a unit vector, and the components of the mean resultant vector are computed as:
\begin{align}
     \bar{x} = \dfrac{1}{N}\sum_{i=1}^{N} \cos \phi_i, 
\qquad
\bar{y} = \dfrac{1}{N}\sum_{i=1}^{N} \sin \phi_i,
    \label{eq:Mean_vector}
\end{align}
from which the mean tidal phase $\bar{\phi}$ is obtained as:
\begin{align}
     \bar{\phi} = \tan^{-1}\!\left(\dfrac{\bar{y}}{\bar{x}}\right)
    \label{eq:Mean_phase}
\end{align}

which identifies a typical phase within the tidal cycle at which events preferentially occur. For example, $\bar{\phi}=0^{\circ}$ indicates that events tend to occur near the peak tidal stress. $\bar{\phi}=-90^{\circ}$ corresponds to events occurring near the peak tidal stress rate.

The resultant vector length $R$ quantifies the concentration of the phase distribution around 
$\bar{\phi}$:
\begin{align}
     R = \sqrt{\bar{x}^2 + \bar{y}^2}, \qquad 0 \leq R \leq 1
    \label{eq:R}
\end{align}

A value of $R=1$ indicates that all events occur at the same tidal phase. In contrast, smaller values of $R$ indicate more scattered distribution of tidal phase, with events distributed over a broader range of phases.

\begin{figure*}
    \centering
    \includegraphics[width=\textwidth]{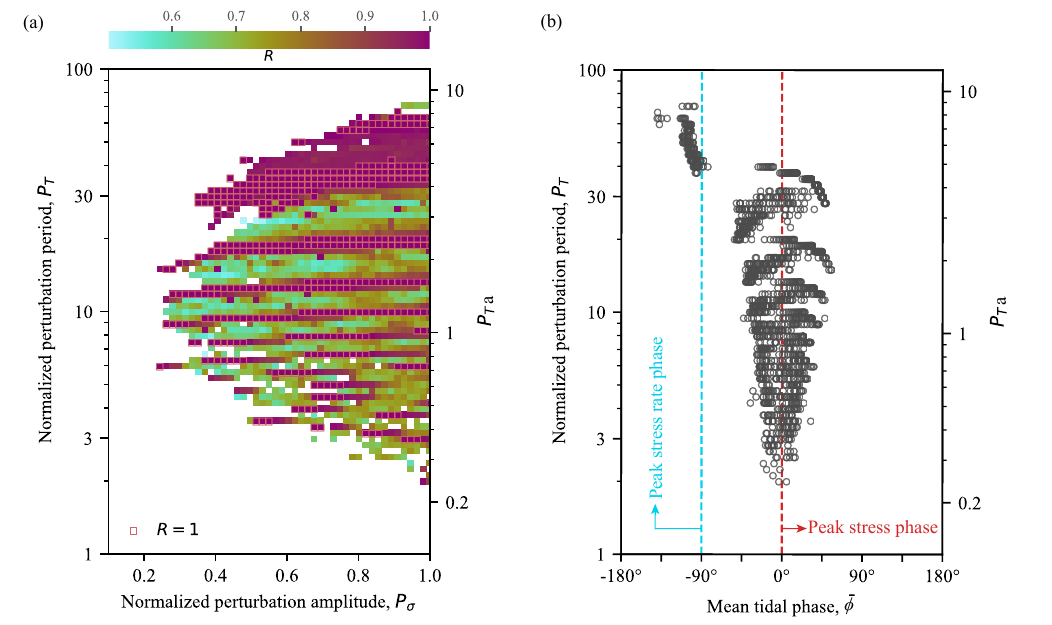}
    \caption{
   Correlation distribution of triggered events. (a) Phase concentration parameter $R$ as a function of normalized perturbation amplitude $P_{\sigma}$ and duration $P_T$, with color indicating the degree of phase clustering (larger $R$ corresponds to stronger concentration around the mean phase). Red squares mark cases with $R=1$. (b) Mean tidal phase $\bar{\theta}$ as a function of $P_T$.  Each circle corresponds to the mean tidal phase calculated from a catalog of triggered events generated by a single simulation run at a given $(P_{\sigma}, P_T)$ parameter combination. Vertical dashed lines indicate the peak stress phase $0^{\circ}$ and the peak stress rate phase $-90^{\circ}$.}
    \label{fig:R_meanphase}
\end{figure*}

\subsubsection{Evidence for resonant triggering}
Figure~\ref{fig:R_meanphase}a shows the tidal phase concentration $R$ of triggered events. Overall, the events exhibit complex yet systematic tidal phase correlations across the explored parameter space. A particularly notable feature is the presence of cases with $R=1$ (red squares), indicating strong phase locking, where events repeatedly occur at the same tidal phase (though not necessarily in every tidal cycle). These phase locking cases form a "comb-like" pattern along the $P_T$ axis that largely coincides with the intervals of enhanced radiation efficiency $\eta_R$ (Figure~\ref{fig:phase_dagram_sbmodel}b), suggesting a close association between phase locking and amplified slip velocity.
Notably, similar "comb-like" amplification is also identified in simulations with boxcar stress perturbation (e.g., Figure~\ref{fig:box_up}), where perturbations with specific perturbation periods produced pronounced slip velocity amplification. Together, these consistent patterns point to a common mechanism: when the perturbation period approaches a resonant timescale of the fault, slip velocity amplification, increased radiation efficiency, and phase locking tend to emerge simultaneously. This consistency provides strong evidence that resonance governs tidal triggering in this regime.

\subsubsection{Phase dependence of triggered events on \texorpdfstring{$P_T$}{PT}}
When examining the mean tidal phase $\bar{\phi}$ of triggered events as a function of $P_T$, a clear systematic trend emerges (Figure~\ref{fig:R_meanphase}b). For relatively large $P_T$ (typically $P_T \gtrsim 40$), $\bar{\phi}$ clusters around $-90^\circ$, indicating preferential occurrence near the phase of peak tidal stress rate.
In contrast, for smaller $P_T$ (approximately $P_T \lesssim 40$), $\bar{\phi}$ shifts toward around $0^\circ$, suggesting a tendency for events to occur closer to the peak tidal stress. This dependence of phase preference on $P_T$ provides a natural framework for interpreting tidal correlations observed in LFEs along the SAF \parencite{van2016fortnightly}. In that study, events cluster near $-90^\circ$ when phases are evaluated using a longer-period (fortnightly) tidal stress envelope, whereas clustering near $0^\circ$ is observed when phases are computed from semidiurnal tidal stresses. Our results further support one interpretation proposed by \textcite{van2016fortnightly} for the apparent $0^\circ$ phase at semidiurnal periods. As illustrated by the time-series examples in Figure~\ref{fig:PTPsigma_vary}b, where small oscillations correspond to individual tidal cycles, in the small $P_T$ regime, events may require multiple perturbation cycles to develop, leading to an apparent alignment with the peak tidal stress rather than with the stress rate. A similar phase dependence was reported by \textcite{bhatnagar2016influence} for microseismicity at the East Pacific Rise, where seismicity preferentially occurred during phases of rapidly increasing fortnightly tidal stress. This behavior is consistent with the large-$P_T$ regime identified in our simulations.

\subsection{Influence of additional model parameters}
 \label{sec:additional_papameters}
All results presented above are obtained using the aging law for state evolution within the RSF framework. As shown in Appendix~\ref{sec:sliplawnondim}, adopting the slip law leads to the same set of governing nondimensional parameters controlling the triggering behavior. We further repeated the same analyses using the slip law. The resulting phase diagrams and relations with perturbations, shown in Figures~\ref{fig:phase_dagram_sbmodel_sliplaw} and \ref{fig:R_meanphase_sliplaw} (~\ref{sec:phase_dagram_sliplaw}), indicate that the overall triggering regime remains broadly similar to that obtained with the aging law, although the quantitative values of $P_T$ and $P_{\sigma}$ associated with different event types are different. The primary difference is that the slip law tends to produce fast events, at smaller $P_{\sigma}$. For example, fast events already emerge around $P_{\sigma}\sim 0.4$ under the slip law, whereas the aging law remains dominated by slower events in the same parameter range. \textcite{Gu1984, ciardoNonlinearStabilityAnalysis2025} also showed that, under the slip law, a VW spring-block system undergoing stable sliding ($k>k_c$) can become unstable when subjected to sufficiently large transient perturbations. In their study, instability is triggered directly by large finite perturbations, such as step changes in load-point motion. In contrast, the present study focuses on relatively small harmonic perturbations. Here, events emerge through resonance-like amplification rather than through direct destabilization by a single large perturbation.

In addition, to account for variable normal stress, we additionally consider the formulation of \textcite{linker1992effects}, in which normal stress variations are coupled to the RSF evolution law. In this formulation, the modified state evolution law is $\dot{\theta}_L = \dot{\theta} - {\alpha\theta}\dot{\sigma}/{b\sigma}$, where $\alpha$ characterizes the sensitivity of the state variable to changes in normal stress. The detailed nondimensional derivation is presented in Appendix~\ref{sec:linkerdieterichnondim}. After nondimensionalization, the parameter $\alpha$ enters the definition of the normalized perturbation amplitude, $P_{\sigma} = (\Delta \tau - (f_*^{ss}-\alpha)\,\Delta \sigma)/(a \sigma_0)$. One extra nondimensional parameter, $\mathcal{L} = \alpha \Delta \sigma / b \sigma_0$, appears in the evolution law, while all other nondimensional parameters remain unchanged. $\mathcal{L}$ represents the variation of the state variable resulting from an transient, small perturbation in normal stress (Eq.~\eqref{eq:theta_step}). In the numerical simulations presented above, we set $\alpha=0$, such that $\mathcal{L}=0$, and the generalized formulation reduces to the case adopted in the Section~\ref{sec:harmonic pertubations}. For laboratory values of $\alpha < f_0$, the parameter $\mathcal{L}$ has range $0 < \mathcal{L}< (a/b) P_\sigma$, and we expect that the phase diagram $(P_T, P_\sigma)$ will remain the same for $\mathcal{L} \ll P_\sigma$. However, for fixed modified $P_{\sigma}$,  increasing $\mathcal{L}$ will lead to more unstable behavior, and slow to fast boundary in phase diagram will move towards left side~(Fig.~\ref{fig:phase_dagram_sbmodel}(a)) . A detailed exploration of the extended $(P_T,P_\sigma,\mathcal{L})$ parameter space is left for future work. 

Moreover, observations of LFEs along the SAF and SSEs in Cascadia show a stronger correlation with relatively small shear stress perturbations than with larger normal stress perturbations \parencite{thomas2009tremor,hawthorneTidalModulationSlow2010}. As discussed by \textcite{hawthorneTidalModulationSlow2010,houston2015low}, one possible explanation for this weak response to normal stress variations is that slip events occurs under undrained conditions, such that changes in normal stress are accompanied by compensating pore pressure variations. Let $h_w$ denote the hydraulic width of the shear zone measured normal to the fault, and let $c_d\,[\mathrm{m^2\,s^{-1}}]$ denote the hydraulic diffusivity. The characteristic diffusion timescale across the shear zone is then $T_d \sim h_w^2/c_d$. For the undrained approximation to be valid, this diffusion timescale must be much longer than both the tidal period, $T_d \gg T$, and the characteristic RSF timescale, $T_d \gg d_c/V_{\rm ss}$~\parencite{segallDilatantStrengtheningMechanism2010}.
If undrained poroelastic effects are included (see Appendix~\ref{sec:poroelastic_effects} for details),  we find that the normalized stress perturbation amplitude can be redefined as
$P_{\sigma} = |\Delta \tau - f_*^{ss}(1-B)\,\Delta \sigma|/(a \sigma_0^\prime)$,
and that the normal stress perturbation parameter becomes
$\epsilon = (1-B)\,\Delta \sigma/\sigma_0^\prime$,
where $B$ is Skempton’s coefficient and $\sigma_0^\prime$ denotes the background effective normal stress. 
In the drained case, additional hydromechanical timescales associated with pore pressure diffusion may become important, potentially introducing additional nondimensional parameters beyond those considered here. Consequently, the analysis presented above remains valid in the presence of undrained poroelastic effects, provided that the nondimensional parameters $P_{\sigma}$ and $\epsilon$ are redefined as above.

In addition, natural tidal stress may involve phase offsets between shear and normal stress perturbations on the fault. As discussed in Appendix~\ref{sec:outofphase_nondim}, such phase differences can be incorporated into the present framework through a modified $$P_\sigma
=
\frac{\sqrt{
\Delta\tau^2
+
(f_*^{ss}\,\Delta\sigma)^2
-
2f_*^{ss}\Delta\tau\Delta\sigma\cos\Delta\phi
}}{a\,\sigma_0}$$ together with an additional nondimensional parameter $${\phi_{\mathrm{eff}}}
=
\tan^{-1}
\left(\frac{
\Delta\tau\sin\Delta\phi} {
\Delta\tau\cos\Delta\phi-f_*^{ss}\Delta\sigma}
\right)$$. For $\Delta\phi=0$, the generalized formulation reduces to the in-phase case adopted in the Section~\ref{sec:harmonic pertubations}. Several observational studies of tremor tidal sensitivity also primarily examine correlations with tidal shear stress alone \parencite{yabe2015tidal}, for which the simplified formulation adopted in the Section~\ref{sec:harmonic pertubations} can still serve as a useful reference single component perturbation model. However, for practical values of $\phi_{\mathrm{eff}}$, the influence of this additional parameter on the phase diagrams has yet to be quantified. A detailed exploration of the extended $(P_T,P_\sigma,\phi_{\mathrm{eff}})$ parameter space is left for the future work. 

We further tested additional parameter combinations to examine the sensitivity of the triggering behavior, including variations in the intrinsic system parameters $R_{ab}$, $\kappa$, and $\mathcal{N}$. While these intrinsic parameters modify the quantitative extent and amplification level of the triggering regimes, the overall triggering structure remains primarily controlled by the perturbation-related parameters $P_T$ and $P_{\sigma}$. In particular, increasing $\kappa$ stabilizes the system and suppresses large velocity amplifications, whereas increasing $R_{ab}$ toward the velocity-neutral regime ($a \approx b$) enhances the susceptibility to resonance-like amplification. The range of $\mathcal{N}=V_{\mathrm{dyn}}/V_{\mathrm{ss}}$ explored here mainly reflects possible variations in the background loading velocity $V_{\mathrm{ss}}$. Although $V_{\mathrm{dyn}}=a\sigma_0/\eta$ is set by the frictional and radiation damping parameters, $V_{\mathrm{ss}}$ may correspond to different loading conditions, from long-term plate  rate to local stressing rates at source patch scale associated with large scale slow slip events. Such differences in boundary conditions can lead to orders-of-magnitude variations in $\mathcal{N}$. Variations in $\mathcal{N}$ mainly affect the overall amplification magnitude while preserving the same qualitative dependence on $P_T$. Importantly, the existence of resonance-like triggering regimes and the systematic transition from stress-controlled to stress-rate-controlled triggering remain robust across all tested cases. Detailed results are presented in~\ref{app:N_kkc_Rab}.

\subsection{Comparison with previous triggering and modulation frameworks}

The previous RSF-based models discussed above differ fundamentally in their unperturbed slip behavior, characteristic controlling timescales, physical triggering or modulation mechanisms, and the fault properties that can be constrained from observations. Table~\ref{tab:comparison_modulation_frameworks} summarizes these major frameworks and their associated physical interpretations.

Among these frameworks, the present study is most closely related to the resonance framework of \textcite{perfettini2001frictional}. We consider a VW patch in the stable sliding regime ($k \gtrsim k_c$), where spontaneous instability does not occur in the absence of external perturbations. Similar to classical seismicity-rate studies, our simulations show a transition between stress-controlled and stress-rate-controlled triggering behavior as the perturbation period varies. However, the physical interpretation is different. Classical seismicity-rate formulations describe the modulation of seismicity rates in already unstable fault and are naturally parameterized by the seismicity relaxation timescale $t_a=a\sigma_0/(kV_{\mathrm{ss}})$. In contrast, the present framework resolves the slip velocity response of an individual stable sliding patch, for which resonance-like amplification is more directly tied to the friction evolution timescale $d_c/V_{\mathrm{ss}}$. We therefore emphasize $P_T=TV_{\mathrm{ss}}/d_c$, while also reporting $P_{Ta}=T/t_a$ to facilitate comparison with seismicity-rate studies. A key result is the identification of a finite triggering window in $(P_T,P_{\sigma})$ space, outside of which harmonic perturbations do not triggered slip events. This differs from classical seismicity-rate formulations, which mainly describe phase and amplitude modulation of event rates rather than the existence or absence of triggering itself. Because $P_T$ explicitly depend on $d_c$, the period dependence identified here may provide constraints on patch dimension.

\begin{table*}[t]
\centering
\caption{Comparison of major RSF-based triggering and modulation frameworks.}
\scriptsize
\begin{tabular}{
p{3.2cm}
p{2.2cm}
p{2.6cm}
p{4.8cm}
p{2.8cm}
}
\hline

Framework / representative studies
&
Unperturbed behavior
&
Key timescale
&
Mechanism
&
primary parameters constraints
\\

\hline

Classical seismicity-rate framework
\parencite{dieterich1994constitutive,heimisson2020analytical}
&
Unstable sliding VW systems  ($k<k_c$)
&
$t_a$
&
harmonic stress perturbation modulates seismicity rates through frictional relaxation controlled by the seismicity response timescale
&
$a\sigma_0$, $t_a$
\\

Hydromechanical seismicity-rate extensions
\parencite{zhao2025tidal}
&
Unstable sliding VW systems ($k<k_c$)
&
$t_a$, $T_f$
&
Dilatancy and pore-fluid diffusion introduce frequency-dependent seismicity modulation
&
$a\sigma_0$, $T_f/t_a$
\\

Creeping fault modulation framework
\parencite{ader2012role,beeler2013inferring}
&
Stable sliding VS systems
&
$T_\theta\sim d_c/V_{\rm ss}$, $T_Q$, $t_a$
&
harmonic stress perturbation amplifies surrounding aseismic creep, which indirectly modulates tremor or LFE activity on embedded patches
&
$a\sigma_0$
\\

Resonance framework
\parencite{perfettini2001frictional}, present study
&
Stable sliding VW patches($k\gtrsim k_c$)
&
$T_c\sim d_c/V_{\rm ss}$
&
Resonance-like amplification can directly trigger slip events when the perturbation period approaches the intrinsic RSF frictional evolution timescale 
&
$a\sigma_0$, $d_c$
\\
\hline
\end{tabular}

\vspace{0.15cm}

\begin{minipage}{\textwidth}
\footnotesize
\textit{Notes:}
$t_a=a\sigma_0/\dot{\tau}_r$ denotes the seismicity relaxation timescale, where $\dot{\tau}_r$ is the background tectonic shear stressing rate;
$T_f=L^2/(4\alpha)$ denotes the characteristic hydraulic diffusion timescale, where $L$ is the characteristic diffusion length scale and $\alpha$ is the hydraulic diffusivity;
$T_\theta=2\pi d_c/V_{\rm ss}$ denotes the RSF frictional evolution timescale;
$T_Q=T_{\theta}a/(a-b)$ denotes the characteristic frictional timescale from the instantaneous direct effect ($a$) to the steady state frictional response ($a-b$).
\end{minipage}

\label{tab:comparison_modulation_frameworks}
\end{table*}

\subsection{Relevance to tidal correlations in natural slow earthquakes}

\subsubsection{Constraints on frictional parameters}  \label{Sec:constrain_friction}
In this study, we have shown that tidal stresses can trigger events on stable sliding VW faults within a specific range of the nondimensional parameters $P_T$ and $P_{\sigma}$ ($P_{\sigma} \gtrsim 0.2$, and  $2 \lesssim P_T \lesssim 70$). A key question is: within what range of structural and frictional parameters do these triggered events actually occur, and are these parameter regimes realistic in natural fault systems?

As discussed above, tidal stress in natural settings reflects the combined contributions of solid Earth tides and ocean tidal loading. It is well established that solid Earth tides typically induce stress changes of
$0.1$--$5\,\mathrm{kPa}$, while ocean tidal loading can generate stress perturbations
reaching up to $\sim 100\,\mathrm{kPa}$ in subduction zones
\parencite{cochran2004earth,zaccagnino2022correlation}.
Taken together, these observations indicate that the effective tidal stress
perturbation generally falls within the range $0.1$--$100\,\mathrm{kPa}$. Here, we do not distinguish between the individual contributions from
shear and normal stress perturbations, but instead consider their combined effect through the quantity $|\Delta \tau - f_*^{ss}\Delta \sigma|$. Combining this estimate with the normalized stress amplitude defined in our framework, $P_{\sigma} = {|\Delta \tau - f_*^{ss} \, \Delta \sigma|}/{a \sigma_0}$, we constrain $a\sigma_0$ to be in the range $0.5$--$500 \,\mathrm{kPa}$, where the lower (upper) bound corresponds to the lower (upper) end of the estimated tidal stress amplitude. This is consistent with the range inferred from tidal correlation analyses of slow earthquakes. As shown in Table~\ref{tab:asigma_constraints}, observational studies involving tidal stress perturbations of order kilopascals (typically $\sim$0.1--a few~kPa) consistently indicate low values of $a\sigma_0$, generally well below the MPa range. These estimates are consistent with the low $a\sigma_0$ values inferred from our modeling framework.

\begin{table}[htbp]
\centering
\small
\caption{A summary of $a\sigma_0$ values constrained from observation studies}
\label{tab:asigma_constraints}
\begin{tabular}{@{}lcccr@{}}
\toprule
\textbf{References} & \textbf{Catalogs} & \textbf{Tidal stress (kPa)} & \textbf{Model} & \textbf{${a \sigma_0}$ (kPa)}  \\
\midrule
\textcite{nakata2008non} & Nankai trough tremors & Coulomb stress $\sim 1$ & \textcite{dieterich1994constitutive} & $1.3$ \\    
\addlinespace

\citeauthor{thomas2009tremor} \citeyear{thomas2009tremor,thomas2012tidal} & Parkfield tremors & shear stress $\sim 0.1$ & \textcite{dieterich1994constitutive} & $0.1{-}1$ \\
\addlinespace

\textcite{beeler2013inferring} & SAF LFEs & shear stress $\sim 0.4$ & \textcite{ader2012role} & $500$ \\
\addlinespace

\textcite{yabe2015tidal} & Nankai \& Cascadia tremors & Coulomb stress $0{-}4$ & \textcite{ader2012role} & $3$ \\
\addlinespace

\textcite{royer2015tidal} & northern Cascadia LFEs & shear stress $\sim 2$ & \textcite{beeler2013inferring} & $7.6$ \\
\addlinespace

\textcite{nakamura2017tidal} & Ryukyu Trench VLFEs & shear stress $\sim 0.4$ & \textcite{beeler2013inferring} & $1{-}2$ \\
\bottomrule
\end{tabular}
\end{table}

We now consider the nondimensional parameter $P_T = {T V_{\mathrm{ss}}}/{d_c}$, where $T$ is the period of a harmonic perturbation. Here, we consider the dominant tidal component with $T = 12 \,\text{hours}$. We assume that the patch is loaded by a background slip velocity $V_{\mathrm{ss}}$, such that the characteristic timescale of the background slip is much longer than both the tidal period and the duration of the triggered slip events. This separation of timescales allows us to treat the background slip velocity as effectively constant during tidal triggering.  We consider two representative cases for $V_{\mathrm{ss}}$. The first case corresponds to the plate convergence rate \parencite{demets2010geologically} or the long-term fault creep rate (the creeping section of the central SAF creeps at about $10^{-9}\,\text{m/s}$ \parencite{thomas2018using}). For this case, $d_c$ is estimated to lie between approximately 0.6 and 20 $\mu$m, consistent with values reported in laboratory rock-friction experiments \parencite{marone1993scaling,marone1998laboratory}. The second case corresponds to local slow slip velocities inferred from ETS, with $V_{\mathrm{ss}}$ in the range $10^{-8}$–$10^{-6}\,\text{m/s}$ \parencite{thomas2018using,rubin2011designer}. For this case, the inferred values of $d_c$ span approximately 6 $\mu$m to 2 cm. Although the upper bound of this estimate is relatively large, similar values have been proposed in the literature; for instance, \textcite{maury2014fault} estimated that a critical slip distance of $5$ cm can reproduce the observed SSE in Mexico. We caution that one must constrain $d_c$ and $a\sigma_0$ values from the phase diagram (for example, from Fig.~\ref{fig:phase_dagram_sbmodel}) carefully, since these values hold only for a fixed $\mathcal{N} = V_{\rm dyn} / V_{\rm ss} = 10^6$, where $V_{\rm ss} = 10^{-9}$ m/s. To apply the main results of this work to different orders of the loading rate $V_{\rm ss}$, such as a larger $V_{\rm ss} = 10^{-6}$ m/s for slow slip events or the extremely low loading rates in intraplate regions $V_{\rm ss} = 10^{-12}$ m/s~\parencite{craigHydrologicallydrivenCrustalStresses2017}), one ideally constructs a new phase diagram for that particular $\mathcal{N}$ value using the code provided with this work~\parencite{ZhouData2026}.

Several previous studies have suggested a possible scale dependence of the characteristic slip distance $d_c$ with patch size. Laboratory experiments linked the characteristic weakening distance to fault surface roughness scales~\parencite{ohnakaConstitutiveScalingLaw2003}, while seismological estimates and recent numerical studies suggest that apparent fracture energy may increase with slip or rupture size~\parencite{abercrombie2005can,gabriel2024fault}. \textcite{ide2005earthquakes} proposed a hierarchical rupture model in which characteristic weakening distance, and hence the fracture energy scales with patch size. Within the RSF framework, \textcite{rubin2005earthquake} showed that fracture energy is proportional to $d_c$ for crack-like ruptures during nucleation. Motivated by these ideas, several studies have explored RSF simulations incorporating size-dependent $d_c$ \parencite{hori2010hierarchical,nakata2023recurrence,gerardi2024geomechanical,almakari2026fault}. Although the physical origin of these possible scaling relationships remains an active topic of investigation, if such relationships are valid, constraints on $d_c$ inferred from sensitivity may also provide indirect insights into the characteristic dimensions of the underlying patches.

\subsubsection{Implications, limitations, and observational consistency} 
In our simulations, the triggered catalogs are more appropriately interpreted as LFE-like events rather than tremors. LFEs are generally understood as repeating failures on a single patch, which is consistent with our modeling framework.
Tremor, in contrast, is often interpreted as the superposition of many such LFEs occurring within a localized region \parencite{shelly2007non,Yabe2025,Shelly2026}. It should be noted, however, that the present simulations consider a single patch in isolation and do not include interactions among multiple patches. As a result, interaction that may be important for tremor generation are not explicitly captured in our model.

In this study, we focus on slip events that are distinct from creeping. Fast and slow events in our simulations are defined purely based on their slip velocities and therefore represent model-based classifications rather than direct equivalents of natural fast or slow earthquakes. It is worth noting, however, that the absence of clear tidal correlations in ordinary earthquake catalogs does not necessarily imply that resonance-like triggering mechanisms are absent. As discussed by \textcite{perfettini2001periodic}, fault systems hosting ordinary earthquake are unlikely to consist of a single homogeneous patch, and small variations in frictional or geometric properties may produce substantial phase dispersion, thereby obscuring coherent tidal correlations at the catalog scale. Similarly, even in our simulations, triggered events may exhibit highly scattered phase distributions despite being generated under harmonic perturbation.

Although the present study primarily focuses on tidal triggering of LFE-like slip events, the resonance-like triggering mechanism explored here may be relevant more broadly to velocity-weakening regions within the brittle-ductile transition. This mechanism may operate across a range of natural fault environments, from creeping segments such as the central creeping section of the SAF to regions characterized by ETS. In such environments, small harmonic stress perturbations may promote slip acceleration on fault patches whose dimensions are comparable to their nucleation size. Depending on the surrounding elastic interactions and fault structure, similar mechanisms may potentially contribute to the modulation of other forms of seismic or aseismic slip, such as tremor activity or repeating earthquakes, under a broader range of transient or harmonic perturbation conditions, including hydrological or seasonal stress perturbations.
 
For example, recent observations by \textcite{zhao2025tidal} showed that ordinary earthquakes at shallower depths on the central SAF are more sensitive to long-period hydrological loading, whereas LFEs respond more strongly to short-period tidal loading. Related studies have also reported significant seasonal modulation of seismicity despite stress amplitudes comparable to semidiurnal solid Earth tides \parencite{sirorattanakul2026seismic}, suggesting that fault systems may respond selectively to external perturbations depending on the perturbation period relative to intrinsic nucleation timescales. Within the framework proposed here, if both populations operate within the triggering regime, their preferred perturbation periods may provide constraints on the corresponding $d_c/V_{\rm ss}$ ratios. This provides a possible interpretation for the scaling of  $d_c$ with patch sizes (also discussed in Sec.~\ref{Sec:constrain_friction}): smaller patches with smaller $d_c$ are expected to respond preferentially to shorter-period perturbations, whereas larger patches with larger $d_c$ may be more sensitive to longer-period loading. More broadly, if  $d_c$ associated with different fault patches varies systematically with patch size or rupture dimension, the observed dependence on perturbation period may also reflect scale-dependent fault responses to perturbation time scales. Exploring these possibilities in continuum fault models remains an important direction for future work.

Several observational studies have reported statistically significant tidal periodicities at subharmonic periods, most commonly near $\sim6$~h and $\sim8$~h, in catalogs of tremors in Taiwan and icequakes on the Ross Ice Shelf, even though these components are weak or nearly absent in the corresponding tidal stress records \parencite{chen2018tidal,yi2025characteristics,udelllopez2026}. Our simulations demonstrate that enhanced radiation efficiency and phase locking emerge only at a discrete set of normalized perturbation periods, forming a "comb-like" pattern (Figures~\ref{fig:phase_dagram_sbmodel}b and~\ref{fig:R_meanphase}a). This indicates that, for a given frictional patch, triggered events can exhibit periodicity only under specific perturbation conditions, rather than across a continuous range of periods. This suggests that, for a given frictional patch, triggered events can occur preferentially at specific perturbation periods rather than across a continuous range, consistent with the isolated subharmonic periodicities reported in observations.

The phase relationship between events and tidal stresses in our simulations depends on $P_T$: for larger $P_T$, events are in phase with the tidal stress rate, while for smaller $P_T$, they are in phase with the tidal stress itself. And it is widely recognized that the tidal correlations of tremors during ETSs are weak in the early stages and become stronger in the later stages, both in the Nankai trough and Cascadia \parencite{houston2015low,yabe2015tidal}. Moreover, \textcite{royer2015tidal} reported that on southern Vancouver Island, the phase relationship between LFEs and tidal stresses evolves from about $-90^\circ$ in the early stage to nearly $0^\circ$ in the later stage. These evolutions can plausibly be attributed to a progressive decrease in the background slip velocity $V_{\mathrm{ss}}$, for example from $10^{-6}$ to $10^{-8}\,\mathrm{m/s}$, corresponding to a decrease in $P_T$ in our model. One discrepancy is that in our simulations the correlation with the tidal stress rate remains high even at large $P_T$, whereas in nature such a clear in-phase relation with the stress rate is not generally observed. This difference may arise because, in nature, the tidal stress rate signal is masked by the much stronger triggering influence of the underlying SSE itself.

A particularly noteworthy finding in our study is that harmonic stress perturbations can even trigger temporally complex slip events on faults that are otherwise stable sliding. This behavior arises from the intrinsic nonlinearity and history dependence of the RSF system. This behavior can be understood from the response to stress step changes. A stress step instantaneously displaces the system away from steady state, either to slip velocities above or below the steady state. Although the system tends to relax back toward steady state in both cases, upward and downward stress steps follow inherently different transient recovery paths, as discussed in Section~\ref{Response to an upward step}.

Although this resonance-based framework is intentionally idealized and natural event catalogs are influenced by multiple interacting processes, it establishes the physical plausibility of resonance as a triggering mechanism and highlights the conditions under which it becomes effective. Accordingly, the nondimensional parameter ranges identified here, together with local tidal characteristics and background slip velocitys, provide a practical framework for interpreting observed tidal correlations and for placing constraints on frictional parameters such as $d_c$ and $a\sigma_0$. Therefore, while isolating resonance effects in natural observations remains challenging, revisiting this mechanism is nevertheless important given its potential role in event triggering processes.

\section{Conclusions}
\label{sec: Conclusions}
Building upon the seminal work of \textcite{perfettini2001frictional}, we identify two key nondimensional parameters, the normalized perturbation period $P_T$ and normalized perturbation amplitude $P_{\sigma}$, controlling tidal triggering in a stable sliding velocity weakening spring-block \rev{system}. Within this framework, resonance-like amplification of slip velocity emerges as a plausible mechanism linking tidal stress to slow earthquake. We further show that the phase relationship depends systematically on the normalized perturbation period $P_T$: for large $P_T$, triggered events occur preferentially near the peak tidal stress rate, whereas for small $P_T$, they align more closely with the peak tidal stress. These results provide a possible explanation for the observed period-dependent sensitivity and phase preference in slow earthquake observations. Finally, comparison between modeled triggering behavior and observed tidal responses suggests that the frictional properties required for triggering under natural fault conditions are physically reasonable.

\section*{Acknowledgments}
YZ, HA, HSB, SI, and AS gratefully acknowledge the support provided through the CNRS-University of Tokyo joint program SESAME. Part of the support was also provided by the European Research Council grant PERSISMO (grant 865411), including the computations conducted on the MADARIAGA HPC cluster at ENS, Paris. AS and HA acknowledge the support of ANR program PREMS (ANR-24-CE56-3575) as well as long standing fruitful discussions on the topic of tidal triggering with T. Hatano, F. Petrelis, K. Chanard and M. Colledge. AG acknowledges the support of LRC Yves Rocard postdoctoral funding from CEA, France. AG and HSB acknowledge the  Indo-French SPARC grant for sponsoring NISER, Bhubaneswar visit which enabled fruitful discussions with Prof. Pathikrit Bhatacharya on modeling harmonic perturbations on rate and state friction faults. AG and HSB also acknowledge discussions with Dr. Jorge Jara from GFZ Potsdam on the role of small perturbations in modulating seismicity. YZ and AG thank Dr. Navid Kheirdast from ISTeP, Paris for fruitful discussions on the energy budget of a event in quasi-dynamic simulations. YZ also thanks Caiyuan Fan, Bharath Shanmugasundaram and Suli Yao from ENS, Paris for helpful discussions. We also used LLM models like ChatGPT and Claude to debug/optimize our codes, and correct grammatical mistakes in the manuscript.

\section*{Open Research}

The datasets generated and analyzed during this study, together with the software and Jupyter notebooks used for the numerical simulations and analysis, are available in a Zenodo repository~\parencite{ZhouData2026}.

\section*{Conflict of Interest}
The authors declare no conflicts of interest relevant to this study.

\printbibliography


\setcounter{section}{0}
\renewcommand\thesection{Appendix \Alph{section}}
\renewcommand\thesubsection{\Alph{section}.\arabic{subsection}}
\renewcommand{\theequation}{\Alph{section}\arabic{equation}}
\setcounter{equation}{0}
\renewcommand{\thefigure}{\Alph{section}\arabic{figure}}
\setcounter{figure}{0}

\makeatletter
\@addtoreset{equation}{section}
\@addtoreset{figure}{section}
\makeatother

\renewcommand{\theHsection}{appendix.\Alph{section}}
\renewcommand{\theHsubsection}{appendix.\Alph{section}.\arabic{subsection}}
\renewcommand{\theHequation}{appendix.\Alph{section}.\arabic{equation}}

\section{Response of a stable sliding RSF fault to transient shear and normal stress perturbations}
\label{app:Instantaneous response of fault to step change}

 We consider an instantaneous change in shear stress from $\tau_1$ to $\tau_1 + \Delta \tau_{\rm step}$, accompanied by a normal stress change from $\sigma_1$ to $\sigma_1 + \Delta \sigma_{\rm step}$,  following~\textcite{paul2024frictional}.

Under RSF with the Linker–Dieterich evolution law~\parencite{linker1992effects}, both the friction coefficient and the state variable undergo instantaneous changes in response to a shear and normal stress step. The friction coefficient responds through the direct effect associated with a sudden change in sliding velocity, while the state variable experiences an instantaneous jump induced by the normal stress perturbation. Immediately before the stress perturbation, the fault is in quasi-static equilibrium and sliding at steady state (${\theta_{\rm ss} V_{\rm ss}}/{d_c} = 1$), such that
\begin{equation}
\tau_1 = f_1 \, \sigma_1,
\end{equation}
where $
f_1 = f(V_{\rm{ss}}, \theta_{\rm ss}) = f_0 + a\log \!\left( {V_{\rm{ss}}}/{V_0} \right)+ b  \log\left({\theta_{\rm ss} V_{0} }/{d_c}\right)$ and $V_{\rm{ss}}$ is the steady-state sliding velocity prior to the perturbation and $\theta_{\rm{ss}}$ is corresponding state variable.

Immediately after the stress step, equilibrium requires
\begin{equation}
\tau_1 + \Delta \tau_{\rm step} = f_2 \left( \sigma_1 + \Delta \sigma_{\rm step} \right),
\end{equation}
where $f_2 = f_0 + a\log\left({V_2}/{V_0} \right)+ b  \log\left({\theta_2 V_0 }/{d_c}\right)$.

Subtracting the pre-step equilibrium condition from the post-step condition yields

\begin{equation}
\Delta \tau_{\rm step} = (f_2 - f_{1})(\sigma_1 + \Delta \sigma_{\rm step}) + f_{1} \Delta \sigma_{\rm step}.
\label{eq:balance_mu}
\end{equation}

During the instantaneous normal stress step, the slip is zero.
Consequently, the state variable does not evolve through slip but changes
according to the Linker-Dieterich normal-stress effect, which can be obtained by solving $$\int_{\theta_{\rm ss}}^{\theta_2} \dfrac{d\theta}{\theta} = - \dfrac{\alpha}{b} \int_{\sigma_1}^{\sigma_1 + \Delta\sigma_{\rm step}} \dfrac{d\sigma}{\sigma},$$ across the normal step,
\begin{equation}
\theta_2=\theta_{\rm ss} \left(\dfrac{\sigma_1+\Delta\sigma_{\mathrm{step}}}{\sigma_1} \right)^{-\dfrac{\alpha}{b}},
\label{eq:theta_step}
\end{equation}
which, for $|\Delta\sigma_{\rm step}| \ll \sigma_1$, reduces to $\theta_2 \approx \theta_{\rm ss}\left(1-\alpha\Delta\sigma_{\rm step}/b\sigma_1\right)$, indicating that the Linker-Dieterich effect produces an instantaneous change in the state variable $\theta$  equal to $\alpha\Delta\sigma_{\rm step}/b\sigma_1$ during a small normal stress step perturbation.

Substituting Eq.~\eqref{eq:theta_step} into RSF framework yields
\begin{equation}
f_2 - f_{1}
=
a \ln\!\left(\dfrac{V_2}{V_{\mathrm{ss}}}\right)
+b\log \left(\dfrac{\sigma_1+\Delta\sigma_{\mathrm{step}}}{\sigma_1} \right)^{-\dfrac{\alpha}{b}}
\label{eq:direct_LD}
\end{equation}
Substituting Eq.~\eqref{eq:direct_LD} into the Eq.~\eqref{eq:balance_mu} yields
\begin{equation}
{V_2}
=V_{\mathrm{ss}} \exp \left(
\dfrac{
\Delta\tau_{\rm step} - f_{1}\,\Delta\sigma_{\rm step}
+
\alpha(\sigma_1 +\Delta\sigma_{\mathrm{step}})\,
\log\!\left(\dfrac{\sigma_1+\Delta\sigma_{\mathrm{step}}}{\sigma_1}\right) 
}{
a\,(\sigma_1+\Delta\sigma_{\mathrm{step}})
}\right).
\label{eq:A7}
\end{equation}
Under the assumption of small perturbations $|\Delta\sigma_{\mathrm{step}}| \ll \sigma_1$,
Eq.~\eqref{eq:A7} reduces to
\begin{equation}
V_2
\approx {V_{\mathrm{ss}}} \exp\left( 
\dfrac{\Delta\tau_{\mathrm{step}}-(f_{1}-\alpha)\,\Delta\sigma_{\mathrm{step}}}{a \,\sigma_1} \right) .\label{eq:linkervelocityjump}
\end{equation}
In the main text, we neglect the normal stress dependence of the state variable
by setting $\alpha = 0$, which simplifies the expression without affecting the
key scaling of the instantaneous velocity response. Also, $\Delta\tau_{\rm step} - (f_1 - \alpha)\,\Delta\sigma_{\rm step} > 0$ means positive Coulomb stress transfer, which will lead to increase in slip velocity, whereas  $\Delta\tau_{\rm step} - (f_1-\alpha)\, \Delta\sigma_{\rm step} < 0$ is negative Coulomb stress transfer leading to decrease in slip velocity of the fault.

\section{Nondimensional equations for the Spring-block model}
\label{app:nondimensional analysics}

Balance equations and the friction law for the quasi-dynamics of a spring-block system :
\begin{align}
    \tau &= \tau_0 + k V_{\rm ss} \,t - k \,\delta - \eta V + \tau_{\mathrm{p}}(t) \label{eq:tau-A} \\
    \sigma &= \sigma_0 + \sigma_{\mathrm{p}}(t) \\
    f(V, \theta) &= f_0 + a \log \left( \dfrac{V}{V_0} \right) + b \log \left( \dfrac{V_0 \theta}{d_c} \right) ,\label{eq:f} \\
    |\tau| &= f(V, \theta) \sigma,
\end{align}
where normal stress perturbation $\sigma_{\mathrm{p}}(t)$ and shear stress perturbation $\tau_{\mathrm{p}}(t)$ induced due to tides are given by:
\begin{align}
\sigma_{\mathrm{p}}(t) = \Delta \sigma \sin \left( 2\pi \dfrac{t}{T} \right) ~~~,~~~ 
\tau_{\mathrm{p}}(t) = \Delta \tau \sin \left( 2\pi \dfrac{t}{T} \right). \label{eq:A_deltatau}
\end{align}

By taking the time derivative of the shear stress balance equation \ref{eq:tau-A}, using ${d \, {|\tau|}}/{d t} = \dot{\tau} \,{\rm sgn}(\tau)$ and assuming ${\rm sgn
}(\tau) > 0$ (no back slip is caused due to harmonic perturbations), we obtain the governing equation for the rate of change of shear stress,
\begin{align}
    \dot{\sigma}f + \dot{f} \sigma &= k (V_{\mathrm{ss}} - V) - \eta \dot{V} + \dot{\tau_{\mathrm{p}}}\nonumber\\ 
    \dot{\sigma}f + \sigma \dfrac{\partial f}{\partial V} \dot{V}+ \sigma \dfrac{\partial f }{\partial \theta } \dot{\theta} 
    &= k (V_{\mathrm{ss}} - V) - \eta \dot{V} + \dot{\tau_{\mathrm{p}}}\nonumber\\
    f\Delta \sigma \dfrac{2\pi}{T}\cos \left( 2\pi \dfrac{t}{T} \right) + \sigma \dfrac{a \dot{V}}{V}+ \sigma \dfrac{b }{\theta }\dot{\theta} 
    &= k (V_{\mathrm{ss}} - V )- \eta \dot{V} + \Delta \tau \dfrac{2\pi}{T}\cos \left( 2\pi \dfrac{t}{T} \right) 
    \label{eq:balanced_bytime}
\end{align}

We introduce $\theta _*=t_* = {d_c}/{V_{\mathrm{ss}}}$ as the state variable timescale, $V_*=V_{\mathrm{ss}} $ as the velocity scale, $f_*^{ss} = f_0 + (a - b) \log \left( {V_{\rm ss}}/{V_0} \right)$ as the friction coefficient scale. Let tildes denote nondimensional quantities.
Thus, $V$, $t$ and $\theta$ in Eq. (\ref{eq:balanced_bytime}) are replaced by their nondimensional forms: $V = V_* \widetilde{V} = V_{\rm ss} \widetilde{V}$, $t = t_* \widetilde{t}$, and $\theta = t_* \widetilde{\theta}$
\begin{align}
 &\dfrac{2\pi\Delta \sigma}{T} \cos \left( 2\pi \dfrac{t_* \widetilde{t}}{T} \right)
\left[ f_0 + a \log \left( \dfrac{V_{\mathrm{ss}}\widetilde{V}}{V_0} \right)
+ b \log \left( \dfrac{V_0\widetilde{\theta} t_*}{d_c} \right) \right] \nonumber 
 + \dfrac{a \sigma \dot{\widetilde{V}}}{t_* \widetilde{V}}
+ \dfrac{b\sigma \dot{\widetilde{\theta}} }{\widetilde{\theta} t_*}\\
&= k V_{\rm ss} (1 - \widetilde{V})
- \dfrac{\eta V_{\rm ss}}{t_*} \dot{\widetilde{{V}}}
+ \Delta \tau \dfrac{2\pi}{T} \cos \left( 2\pi \dfrac{t_* \widetilde{t}}{T} \right)
\end{align}

Multiplying each side of the equation by $t_*/(a\sigma_0)$ and substituting
\begin{align*}
f_*^{ss} &= f_0 + (a-b)\log(V_{\rm ss}/V_0), \\
k_c      &= (b-a)\sigma_0/d_c, \\
V_{\rm dyn} &= a\sigma_0/\eta,
\end{align*}
we get
\begin{align}
&\dfrac{\Delta \sigma}{\sigma_0} \dfrac{2\pi t_*}{T} \cos \left( 2\pi \dfrac{t_* \widetilde{t}}{T} \right) \log \widetilde{V}
+ \dfrac{b \Delta \sigma}{a\sigma_0} \dfrac{2\pi t_*}{T} \cos \left( 2\pi \dfrac{t_* \widetilde{t}}{T} \right) \log \widetilde{\theta}  \nonumber + \dfrac{\sigma \dot{\widetilde{V}}}{ \sigma_0 \widetilde{V} }
+ \dfrac{b\sigma \dot{\widetilde{\theta}} }{a \sigma_0\widetilde{\theta} }\\
&= \dfrac{k}{k_c}\left(\dfrac{b-a}{a}\right)(1 - \widetilde{V})
- \dfrac{ V_{\rm ss}}{V_{\rm dyn}} \dot{\widetilde{V}}
+ \left(\dfrac{\Delta \tau - \Delta \sigma f_*^{ss}}{a\sigma_0} \right)\dfrac{2\pi t_*}{T} \cos \left( 2\pi \dfrac{t_* \widetilde{t}}{T} \right)
\label{eq:b7}
\end{align}

We can now clearly identify six nondimensional parameters: $P_T = T/t_*$ (normalized period), $P_{\sigma} = |\Delta \tau - f_*^{ss} \Delta \sigma|/(a \sigma_0)$ (normalized stress perturbation amplitude), $\epsilon = \Delta \sigma/\sigma_0$ (normalized normal stress perturbation), $\kappa = k/k_c$ (normalized stiffness), $R_{ab} = a/b$ (RSF parameter), and $\mathcal{N} = V_{\rm dyn}/V_{\mathrm{ss}}$ (normalized radiation damping). Rewriting the above equation in terms of these parameters, we obtain following nondimensional equation:
\begin{align}
\dfrac{2\pi}{\boldsymbol{P_T}} \cos\left( \dfrac{2\pi\widetilde{t}}{\boldsymbol{P_T}} \right) &\left(\boldsymbol{\epsilon}\log \widetilde{V} + \dfrac{\boldsymbol{\epsilon}}{\boldsymbol{R_{ab}}} \log \widetilde{\theta} - \boldsymbol{P_{\sigma}}\right) + \dfrac{\dot{\widetilde{V}}}{\widetilde{V}} \left[1 + \boldsymbol{\epsilon} \sin\left( \dfrac{2\pi\widetilde{t}}{\boldsymbol{P_T}} \right) + \dfrac{\widetilde{V}}{\boldsymbol{\mathcal{N}}}\right] \nonumber\\
&\quad + \dfrac{\dot{\widetilde{\theta}}}{\boldsymbol{R_{ab}}\widetilde{\theta}} \left[1 + \boldsymbol{\epsilon} \sin\left( \dfrac{2\pi\widetilde{t}}{\boldsymbol{P_T}} \right)\right] = \boldsymbol{\kappa}\left(\dfrac{1-\boldsymbol{R_{ab}}}{\boldsymbol{R_{ab}}}\right) (1 - \widetilde{V}),
\label{eq:non_demensional_general}
\end{align}
with six nondimensional parameters $\boldsymbol{R_{ab}}$, $\boldsymbol{\kappa}$, $\boldsymbol{\mathcal{N}}$, $\boldsymbol{\epsilon}$, $\boldsymbol{P_T}$, and $\boldsymbol{P_\sigma}$.

The nondimensional tidal perturbation amplitude $P_{\sigma} = |\Delta\tau-f_*^{ss}\Delta\sigma|/{a \sigma_0}$ characterizes this instantaneous response of the fault due to tidal loading. However, due to the persistent nature of the tidal loading, the sign of the Coulomb stress $\Delta\tau-f_*^{ss}\Delta\sigma$ for the tidal loading doesn't affect the long-term dynamics of the model and hence the absolute sign of it is taken while defining this nondimensional parameter.

\subsection{Aging law}

The evolution of the nondimensional state variable as governed by the aging law,
\begin{equation}
\dot{\widetilde{\theta}} = 1- \widetilde{V} \widetilde{\theta} 
\end{equation}
Now using Eq.~\eqref{eq:non_demensional_general}, substituting the above definition of evolution aging law, we get:
\begin{align}
\dfrac{2\pi}{\boldsymbol{P_T}} \cos\left( \dfrac{2\pi\widetilde{t}}{\boldsymbol{P_T}} \right) &\left(\boldsymbol{\epsilon}\log \widetilde{V} + \dfrac{\boldsymbol{\epsilon}}{\boldsymbol{R_{ab}}} \log \widetilde{\theta} - \boldsymbol{P_{\sigma}}\right) + \dfrac{\dot{\widetilde{V}}}{\widetilde{V}} \left[1 + \boldsymbol{\epsilon} \sin\left( \dfrac{2\pi\widetilde{t}}{\boldsymbol{P_T}} \right) + \dfrac{\widetilde{V}}{\boldsymbol{\mathcal{N}}}\right] \nonumber\\
&\quad + \dfrac{1}{\boldsymbol{R_{ab}}\widetilde{\theta}} \left(1-\widetilde{V} \widetilde{\theta}\right) \left[1 + \boldsymbol{\epsilon} \sin\left( \dfrac{2\pi\widetilde{t}}{\boldsymbol{P_T}} \right)\right] = \boldsymbol{\kappa}\left(\dfrac{1-\boldsymbol{R_{ab}}}{\boldsymbol{R_{ab}}}\right) (1 - \widetilde{V})
\end{align}

When we have only shear stress perturbation and $\epsilon \rightarrow 0$,
\begin{align}
\dot{\widetilde{V}} = \dfrac{\widetilde{V}}{1 + {\widetilde{V}}/{\boldsymbol{\mathcal{N}}}} \left[\boldsymbol{\kappa}\left(\dfrac{1-\boldsymbol{R_{ab}}}{\boldsymbol{R_{ab}}}\right) (1 - \widetilde{V}) - \dfrac{1}{\boldsymbol{R_{ab}}\widetilde{\theta}} \left(1-\widetilde{V} \widetilde{\theta}\right) + \boldsymbol{P_{\sigma}}\, \dfrac{2\pi }{\boldsymbol{P_T}} \cos\left( \dfrac{2\pi\widetilde{t}}{\boldsymbol{P_T}} \right)   
\right],
\end{align}

When there is no stress perturbation, i.e., $\boldsymbol{\epsilon} = 0$ \& $\boldsymbol{P_{\sigma}} = 0$, the above equation reduces to:
\begin{align}
\dot{\widetilde{V}} = \dfrac{\widetilde{V}}{1 + {\widetilde{V}}/{\boldsymbol{\mathcal{N}}}} \left[\boldsymbol{\kappa}\left(\dfrac{1-\boldsymbol{R_{ab}}}{\boldsymbol{R_{ab}}}\right) (1 - \widetilde{V}) - \dfrac{1}{\boldsymbol{R_{ab}}\widetilde{\theta}} \left(1-\widetilde{V} \widetilde{\theta}\right)\right],
\end{align}
with three nondimensional numbers $\boldsymbol{R_{ab}}$, $\boldsymbol{\kappa}$, and $\boldsymbol{\mathcal{N}}$.

\subsection{Slip law} \label{sec:sliplawnondim}
The evolution of the nondimensional state variable as governed by the slip law,
\begin{equation}
\dot{\widetilde{\theta}} = - \widetilde{V} \widetilde{\theta} \log(\widetilde{V} \widetilde{\theta})
\end{equation}
Now using Eq.~\eqref{eq:non_demensional_general}, substituting above definition of evolution slip law, we get
\begin{align}
\dfrac{2\pi}{\boldsymbol{P_T}} \cos\left( \dfrac{2\pi\widetilde{t}}{\boldsymbol{P_T}} \right) &\left(\boldsymbol{\epsilon}\log \widetilde{V} + \dfrac{\boldsymbol{\epsilon}}{\boldsymbol{R_{ab}}} \log \widetilde{\theta} - \boldsymbol{P_{\sigma}}\right) + \dfrac{\dot{\widetilde{V}}}{\widetilde{V}} \left[1 + \boldsymbol{\epsilon} \sin\left( \dfrac{2\pi\widetilde{t}}{\boldsymbol{P_T}} \right) + \dfrac{\widetilde{V}}{\boldsymbol{\mathcal{N}}}\right] \nonumber\\
&\quad - \dfrac{1}{\boldsymbol{R_{ab}}} \widetilde{V} \log(\widetilde{V} \widetilde{\theta}) \left[1 + \boldsymbol{\epsilon} \sin\left( \dfrac{2\pi\widetilde{t}}{\boldsymbol{P_T}} \right)\right] = \boldsymbol{\kappa}\left(\dfrac{1-\boldsymbol{R_{ab}}}{\boldsymbol{R_{ab}}}\right) (1 - \widetilde{V})
\end{align}
Note that only one term is modified, but it has no impact on the nondimensional parameters.

\subsection{Linker-Dieterich evolution law} \label{sec:linkerdieterichnondim}
When we consider the Linker and Dieterich evolution law~\parencite{linker1992effects}:
\begin{align}
    \dot{\theta_L} &= 1 - \dfrac{V \theta}{d_c}-\dfrac{\alpha}{b}\,\theta \dfrac{\dot{\sigma}}{\sigma}
    \label{eq:linker_evolution}
\end{align}

The evolution of the nondimensional state variable as follows:
\begin{equation}
\dot{\widetilde{\theta}}
= 1 - \widetilde{V}\, \widetilde{\theta}
- \boldsymbol{\mathcal{L}} \,\widetilde{\theta}\,\dfrac{2 \pi }{ \boldsymbol{P_T}\left[1+\boldsymbol{\epsilon}\sin\!\left(\dfrac{2\pi \widetilde{t}}{\boldsymbol{P_T}}\right)\right]}
\cos\!\left(\dfrac{2\pi \widetilde{t}}{\boldsymbol{P_T}}\right),
\label{eq:linker_diterich}
\end{equation}
 where $\boldsymbol{\mathcal{L}} ={\alpha \Delta \sigma}/{b\sigma_0}$. 
Now using Eq.~\eqref{eq:non_demensional_general}, substituting Equation~\eqref{eq:linker_evolution}, we get:
\begin{equation}
\begin{aligned}
\dfrac{2\pi}{\boldsymbol{P_T}} \cos\left( \dfrac{2\pi\widetilde{t}}{\boldsymbol{P_T}} \right) &\left(\boldsymbol{\epsilon}\log \widetilde{V} + \dfrac{\boldsymbol{\epsilon}}{\boldsymbol{R_{ab}}} \log \widetilde{\theta} - \boldsymbol{P_{\sigma}}\right) + \dfrac{\dot{\widetilde{V}}}{\widetilde{V}} \left[1 + \boldsymbol{\epsilon} \sin\left( \dfrac{2\pi\widetilde{t}}{\boldsymbol{P_T}} \right) + \dfrac{\widetilde{V}}{\boldsymbol{\mathcal{N}}}\right] \\
&\quad + \dfrac{1}{\boldsymbol{R_{ab}}\widetilde{\theta}}
\left[1 - \widetilde{V}\, \widetilde{\theta}
- \dfrac{2 \pi \alpha \boldsymbol{\epsilon} \widetilde{\theta}}{ b\boldsymbol{P_T}\left[1+\boldsymbol{\epsilon}\sin\!\left(\dfrac{2\pi \widetilde{t}}{\boldsymbol{P_T}}\right)\right]}
\cos\!\left(\dfrac{2\pi \widetilde{t}}{\boldsymbol{P_T}}\right)\right]\left[1 + \boldsymbol{\epsilon} \sin\left( \dfrac{2\pi\widetilde{t}}{\boldsymbol{P_T}} \right)\right] \\
&= \boldsymbol{\kappa}\left(\dfrac{1-\boldsymbol{R_{ab}}}{\boldsymbol{R_{ab}}}\right) (1 - \widetilde{V}),
\end{aligned}
\end{equation}

Rearranging the above equation, we obtain:
\begin{align}
\dfrac{2\pi}{\boldsymbol{P_T}} \cos\left( \dfrac{2\pi\widetilde{t}}{\boldsymbol{P_T}} \right) &\left(\boldsymbol{\epsilon}\log \widetilde{V} + \dfrac{\boldsymbol{\epsilon}}{\boldsymbol{R_{ab}}} \log \widetilde{\theta} - \boldsymbol{P_{\sigma}} \right) 
+ \dfrac{\dot{\widetilde{V}}}{\widetilde{V}} \left[1 + \boldsymbol{\epsilon} \sin\left( \dfrac{2\pi\widetilde{t}}{\boldsymbol{P_T}} \right) + \dfrac{\widetilde{V}}{\boldsymbol{\mathcal{N}}}\right] \nonumber\\
&\quad + \dfrac{1}{\boldsymbol{R_{ab}}\widetilde{\theta}}
\left(1 - \widetilde{V}\, \widetilde{\theta}\right)
\left[1 + \boldsymbol{\epsilon} \sin\left( \dfrac{2\pi\widetilde{t}}{\boldsymbol{P_T}} \right)\right] 
= \boldsymbol{\kappa}\left(\dfrac{1-\boldsymbol{R_{ab}}}{\boldsymbol{R_{ab}}}\right) (1 - \widetilde{V})
\end{align}

where we can new definition of $\boldsymbol{P_{\sigma}} = {[\Delta \tau - (f_*^{ss}-\alpha) \, \Delta \sigma]}/{a \sigma_0}$ in terms of effective friction coefficient $f_{ss}^* - \alpha$. As pointed out by \textcite{rice2001a}, and as can be observed in Eq.~\eqref{eq:linkervelocityjump}, $(f_*^{ss}-\alpha)$ acts as an effective friction coefficient during an instantaneous normal stress change, which comes automatically from above nondimensionalization.

Note that the nondimensionalized equilibrium equations remain similar and still contain six nondimensional numbers. However, the evolution law introduces one additional nondimensional parameter, $\boldsymbol{\mathcal{L}}={\alpha \Delta \sigma}/{b\sigma_0}$, which, in the small-perturbation limit of Eq.~\eqref{eq:theta_step}, represents the instantaneous change of the state variable $\theta$,  due to small normal stress perturbations. Therefore the parametric space with Linker-Dieterich evolution law has seven nondimensional parameters $\boldsymbol{R_{ab}}$, $\boldsymbol{\kappa}$, $\boldsymbol{\mathcal{N}}$, $\boldsymbol{\epsilon}$, $\boldsymbol{P_T}$, $\boldsymbol{P_\sigma}$ and $\boldsymbol{\mathcal{L}}$.

\subsection{Consideration of poroelastic effects in undrained limit}
\label{sec:poroelastic_effects}
In this undrained limit, the effective normal stress, $\sigma^\prime$ can be written as
\begin{equation}
\sigma^\prime =\sigma-p= \sigma_0 + \sigma_{\mathrm{p}}(t) - (p_0 + \Delta p(t)) = \sigma_0  - p_0 + (1-B)\sigma_{\mathrm{p}}(t) = \sigma_0^\prime + (1-B) \sigma_{\mathrm{p}}(t),
\end{equation}
where $\sigma$ denotes the total normal stress, decomposed into a background component $\sigma_0$ and a time-dependent perturbation $\sigma_{\mathrm{p}}(t)$, and $p$ is the pore fluid pressure, consisting of an initial pore pressure $p_0$ and a perturbation $\Delta p(t)$. The quantity $\sigma_0^\prime = \sigma_0  - p_0$ represents the initial effective normal stress. Under undrained conditions, the pore pressure change induced by the normal stress perturbation satisfies $\Delta p(t) = B \sigma_{\mathrm{p}}(t) $, where $B$ is the Skempton's pore pressure coefficient, defined as the ratio of induced pore pressure change to the applied confining stress change \parencite[Eq.~10.1]{segallEarthquakeVolcanoDeformation2010}.

Now using Eq.~\eqref{eq:eq_sb} of the main text, substituting above definition of effective normal stress and and nondimensionalizing, we get

\begin{align}
\dfrac{2\pi}{\boldsymbol{P_T}} \cos\left( \dfrac{2\pi\widetilde{t}}{\boldsymbol{P_T}} \right) &\left(\boldsymbol{\epsilon}  \log \widetilde{V} + \dfrac{\boldsymbol{\epsilon} }{\boldsymbol{R_{ab}}} \log \widetilde{\theta} - \boldsymbol{P_{\sigma}}\right) + \dfrac{\dot{\widetilde{V}}}{\widetilde{V}} \left[1 + \boldsymbol{\epsilon} \sin\left( \dfrac{2\pi\widetilde{t}}{\boldsymbol{P_T}} \right) + \dfrac{\widetilde{V}}{\boldsymbol{\mathcal{N}}}\right] \nonumber\\
&\quad + \dfrac{1}{\boldsymbol{R_{ab}}\widetilde{\theta}} \left(1-\widetilde{V} \widetilde{\theta}\right) \left[1 + \boldsymbol{\epsilon} \sin\left( \dfrac{2\pi\widetilde{t}}{\boldsymbol{P_T}} \right)\right] = \boldsymbol{\kappa}\left(\dfrac{1-\boldsymbol{R_{ab}}}{\boldsymbol{R_{ab}}}\right) (1 - \widetilde{V})
\end{align}

where
 $\boldsymbol{P_{\sigma}} = {|\Delta \tau - f_*^{ss}(1-B) \, \Delta \sigma|}/{a \sigma_0^\prime}$, and
 $\boldsymbol{\epsilon} ={(1-B) \, \Delta \sigma}/{\sigma_0^\prime}$, which gives exactly same nondimensional equation as shown in main text. Therefore, entire analysis of main text holds for this case with new definition of nondimensional parameters $P_\sigma$ and $\epsilon$.

\subsection{Out-of-phase shear and normal stress perturbations}
\label{sec:outofphase_nondim}

The above derivation assumes that the shear and normal stress perturbations are in phase. 
A more general case can be written by introducing a constant phase offset $\Delta\phi$ between the two components:
\begin{align}
\sigma_{\mathrm{p}}(t) &= \Delta \sigma \sin \left( 2\pi \dfrac{t}{T} \right)~~~;~~~ 
\tau_{\mathrm{p}}(t) = \Delta \tau \sin \left( 2\pi \dfrac{t}{T}+\Delta\phi \right).
\end{align}

In this case, the last term in Eq.~\eqref{eq:b7} becomes
\begin{align}
\dfrac{2\pi t_*}{T}
\left[
\dfrac{\Delta\tau}{a\sigma_0}
\cos\left(\dfrac{2\pi t_*\widetilde{t}}{T}+\Delta\phi\right)
-
\dfrac{f_*^{ss}\Delta\sigma}{a\sigma_0}
\cos\left(\dfrac{2\pi t_*\widetilde{t}}{T}\right)
\right].
\end{align}
This can be rewritten as
\begin{align}
 \boldsymbol{P_{\sigma}}\dfrac{2\pi}{ \boldsymbol{P_T}}
\cos\left(\dfrac{2\pi\widetilde{t}}{ \boldsymbol{P_T}}+{ \boldsymbol{\phi_\mathrm{eff}}}\right),
\end{align}
where $${P_\sigma}
=
\frac{
\sqrt{
\Delta\tau^2
+
(f_*^{ss}\Delta\sigma)^2
-
2f_*^{ss}\Delta\tau\Delta\sigma\cos\Delta\phi
}}{a\,\sigma_0},$$ and 
$${\phi_{\mathrm{eff}}}
=
\tan^{-1}
\left[\frac{
\Delta\tau\sin\Delta\phi}{
\Delta\tau\cos\Delta\phi-f_*^{ss}\Delta\sigma}
\right].$$ Here, $\phi_{\mathrm{eff}}$ is the effective phase of the combined Coulomb perturbation, represents the effective phase shift of the combined perturbation term relative to the normal stress perturbation. Thus, when shear and normal stress perturbations are out of phase, the governing equations retain the same overall nondimensional structure, but with the modified $P_\sigma$ and an additional nondimensional parameter $\phi_{\mathrm{eff}}$. For $\Delta\phi=0$, the generalized expression reduces to the definition used in the main text.

\section{Radiation efficiency of a quasi-dynamic spring-block system}
\label{app:eta}

While the governing quasi-dynamic equation for the spring-block system Eq.~\eqref{eq:eq_sb} holds at every instant of the time, our primary interest lies in the energy changes accumulated over a finite time window $[t_1,t_2]$, which is identified as a single event based on a velocity threshold criterion. The threshold is chosen such that $V(t_1), V(t_2) \ll V_{\rm dyn} =  {a \sigma_0}/{\eta}$ so that we attain quasi-static equilibrium at end of time windows of an event. To analyze the energy budget of such an event, we subtract the quasi-dynamic governing equation evaluated at time $t_2$ from the Eq.~\eqref{eq:eq_sb}. This yields an incremental force balance relative to the window endpoint,
\begin{equation}
k \left[ (V_{\rm ss} t - \delta(t)) - (V_{\rm ss} t_2 - \delta_2) \right]
=
(\tau(t) - \tau_{2})+
\eta\, (V(t) - V_2)
-
(\tau_{\mathrm{p}}(t) - \tau_{p2}),
\label{eq:incremental_force}
\end{equation}
where $\delta_2=\delta(t_2), \tau_{p2} = \tau_{p}(t_2), \tau_{2} = \tau(t_2) $ and $V_2=V(t_2)$.
The term on the left side in Eq.~\eqref{eq:incremental_force} can be rewritten as
\begin{equation}
k \left[ (V_{\rm ss} t - \delta(t)) - (V_{\rm ss} t_2 - \delta_2) \right]
=
k(\delta_2-\delta(t))
+
k V_{\rm ss}(t-t_2).
\end{equation}
Multiplying Eq.~\eqref{eq:incremental_force} by the slip velocity $V(t)$ gives the incremental power balance,
\begin{equation}
k(\delta_2-\delta) V
+
k V_{\mathrm{ss}}(t-t_2)V
=
(\tau-\tau_{2})V
+
\eta (V-V_2)V
-
(\tau_{\mathrm{p}}-\tau_{p2})V .
\label{eq:incremental_power}
\end{equation}
Integrating Eq.~\eqref{eq:incremental_power} over the event duration $[t_1,t_2]$ yields the incremental energy balance. The first spring-related term can be evaluated as
\begin{equation}
\int_{t_1}^{t_2} k(\delta_2-\delta)\,V\,dt
=
\dfrac{1}{2}k(\delta_2-\delta_1)^2,
\end{equation}
where $\delta_1=\delta(t_1)$.

The second spring-related term involves the background loading and is evaluated using integration by parts,
\begin{equation}
\int_{t_1}^{t_2} k V_{\rm ss}(t-t_2)\,V\,dt
=
-\,k V_{\rm ss} \int_{t_1}^{t_2} (\delta-\delta_1)\,dt .
\end{equation}

Similarly, collecting other terms and using integration by parts, the incremental energy balance over the event window can be written as
\begin{align}
-\, \int_{t_1}^{t_2} (\dot{\tau_{\mathrm{p}}} + k V_{\mathrm{ss}}) (\delta-\delta_1)\,dt
+
\dfrac{1}{2}\,k\,(\delta_2-\delta_1)^2
&=
\int_{t_1}^{t_2} (\tau-\tau_{2})V\,dt
+
\int_{t_1}^{t_2} \eta (V-V_2)V\,dt
.
\label{eq:incremental_energy}, \\
-\Delta W_{\rm ext} + \Delta W_o &= D_o + E_R.
\end{align}
Equation~\eqref{eq:incremental_energy} represents the energy budget of a velocity-threshold-defined event, expressed in terms of energy increments relative to the window endpoint $t_2$. This formulation follows directly from quasi-dynamic governing equation and no assumption are made to obtain the above equation. Above equation can be compared to \textcite[Eq.~2.24]{kostrovSeismicMomentEnergy1974}. The main feature of this equation is that all the terms can be calculated using incremental quantities. 

Assuming $A$ $\rm [m^2]$ is the rupture area of the event, the last term in Eq.~\eqref{eq:incremental_energy} corresponds to the radiated energy $\rm [Nm]$ for quasi-dynamic approximation, $$E_R = A \int_{t_1}^{t_2} \eta \,(V-V_2)V\,dt \approx A \int_{t_1}^{t_2} \eta \,V^2 \,dt $$~\parencite[Eq.~7]{senatorskiRadiatedEnergyEstimations2014}, if we assume at  $t_2$ the fault satisfy static equilibrium or $V_2 \ll V(t)\,\, \forall\, t \in (t_1, t_2)$. \textcite[Eq.~2.1]{kostrovSeismicMomentEnergy1974} defined the radiated energy as additional work done by a seismic source to its surrounding when the source is active. This energy is the one that is felt in terms of seismic waves on Earth’s surface and is present irrespective of speed of the event. The radiated energy in principle can be calculated from spectra for earthquakes ~\parencite{abercrombie2005can}. 

The first term on the right side is an observable energy  dissipated due to the frictional work $\rm [Nm]$, $D_o = A \int_{t_1}^{t_2} (\tau-\tau_{2})V\,dt = - A \int_{t_1}^{t_2} \dot{\tau} (\delta - \delta_1)\,dt $. It is not possible to calculate this quantity directly from observations unless we perform numerical modeling. The second term on the left side is an observable elastic strain energy $\rm [Nm]$, released during the event, $\Delta W_o = A \dfrac{1}{2}\,k\, (\delta_2 - \delta_1)^2$, which depends on the static stress drop, $k (\delta_2 - \delta_1)$, and the slip during the event ($\delta_2 - \delta_1$), all  quantifiable in observations. Note that total strain energy release $[\rm Nm]$ during  an event is $\Delta W_T = A \dfrac{1}{2} (\tau_{1} + \tau_{2})(\delta_2 - \delta_1)$ ~\parencite[Box 3.4]{aki2002} (neglecting the contribution due to the external loading during the event), however it cannot be directly calculated by observations as it depends on average shear stress during the event.

 \textcite{kostrovSeismicMomentEnergy1974} originally derived the energy budget equation for ordinary earthquakes with typical duration a few seconds/minutes, and the work done by the external perturbations and tectonic loading $\rm [Nm]$, $\Delta W_{\rm ext} = A \int_{t_1}^{t_2} (\dot{\tau_{\mathrm{p}}} + k V_{\mathrm{ss}}) (\delta-\delta_1)\,dt$, during the event is assumed zero in his derivation. For SSEs, which can have duration of days/months, the external work done can be non-negligible. However, it will be difficult to constrain the exact details of loading in natural observations. Therefore, we write above incremental energy balance as
\begin{align}
    \Delta W_o &=  D_o + E_R + \Delta W_{\rm ext},  \nonumber\\
    &\approx D_o + E_R.
\end{align}

\textcite[Ch.~3.2]{venkataramanInvestigatingMechanicsEarthquakes2002}
 and \textcite{kanamoriMicroscopicMacroscopicPhysics2000} recommend a working definition of radiation efficiency in terms of observable quantities,
\begin{equation}
    \eta_R := \dfrac{E_R}{\Delta W_o} \approx  \dfrac{E_R}{D_o + E_R},
\end{equation}
which is calculated in our numerical simulations for all the events and average $\eta_R$ of all events for particular parameters ($P_T$, $P_\sigma$) is shown in Figure~\ref{fig:phase_dagram_sbmodel}b.  In our numerical simulations, all the  terms in the  Equation~\eqref{eq:incremental_energy} are computed by defining the path dependent integrands of $D_o$, $E_R$, and $\Delta W_{\rm ext}$ as state variables of the ODE system so that the running energy balance can be computed in postprocessing between any interval $t_1$ and $t_2$. We explicitly verify the validity of the incremental energy balance for each events defined by velocity thresholds and examine the external work term $\Delta W_{\mathrm{ext}}$ and find that its contribution is negligible in our simulations.

\section{Numerical Methods}
\label{sec:Numerical Methods}
We solved the governing equations using the \texttt{DifferentialEquations.jl} package in Julia 
(v1.11.0, DifferentialEquations.jl v7.15.0). Time integration was performed with an adaptive explicit Runge--Kutta method (\texttt{ExplicitRK}), using the Cash-Karp Butcher table (\texttt{tableau=constructCashKarp}) \parencite{cash1990}.

The solver was configured with an absolute tolerance of $\texttt{abstol} = 10^{-14}$ and a relative tolerance of $\texttt{reltol} = 10^{-12}$, 
together with a maximum iteration limit of $\texttt{maxiters} = 10^9$. 
Error control was applied only to the physically relevant state variables $(\tau, V, \sigma)$, excluding auxiliary integral variables.
The integration interval was set to $t \in [0, \, 20000 \, t_\ast]$, where $t_\ast = d_c / V_{\mathrm{ss}}$. 

In a few simulations, we observe that long sequences of slip events are followed by a transition to persistent stable oscillations. Upon verification, these late-time oscillations are attributed to numerical error accumulation, which cannot be eliminated by further increasing the time-step tolerance. Since such cases are rare, we control for them by requiring a minimum number of slip events so that our overall conclusions remain unaffected.

\section{Analysis of slip law}
\label{sec:phase_dagram_sliplaw}
Using the same set of nondimensional parameters as in the aging law case, we find that the overall trends remain consistent for slip law. Triggered events occur only for a limited range of normalized loading periods and when the normalized perturbation amplitude is sufficiently large. However, for identical parameter values, the slip law tends to produce events with systematically higher values of $\eta_R$, indicating a greater propensity for high-$\eta_R$ triggering.

\begin{figure}
    \centering
    \includegraphics[width=\linewidth]{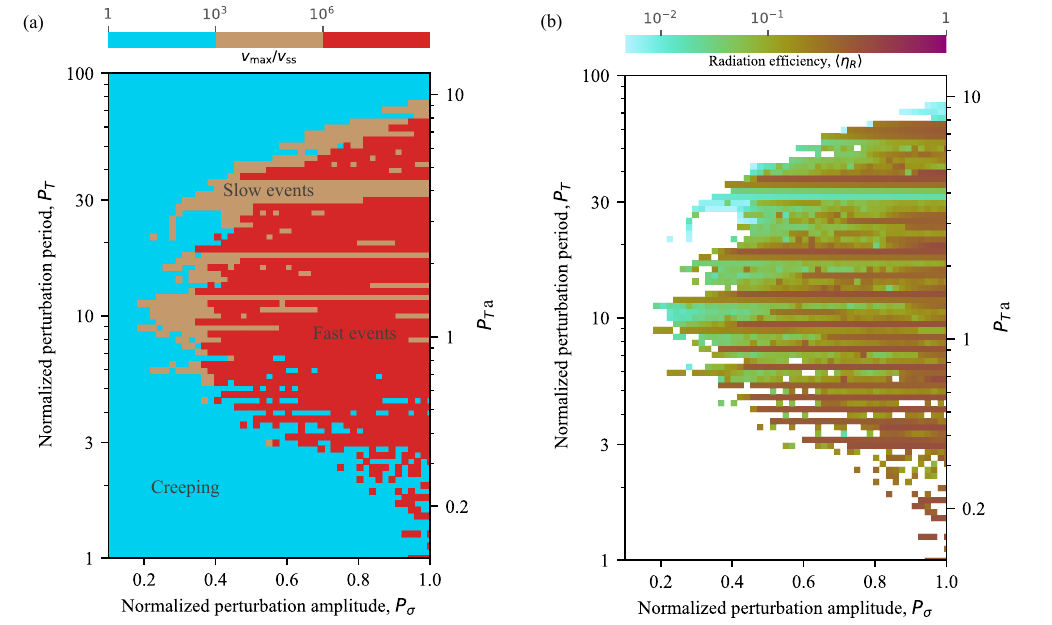}
    \caption{Response phase diagrams of a stable sliding VW fault in the $(P_{\sigma}, P_T)$ parameter space under harmonic normal stress perturbations for slip law. (a) Phase diagram colored by the normalized maximum slip velocity $V_{\max}/V_{\mathrm{ss}}$. Light blue indicates creeping behavior, brown denotes slow slip events, and red corresponds to fast events.(b) The same parameter space colored by the average radiation efficiency $\langle \eta_R \rangle$, computed only for cases where slip events occur. An event is defined as a slip episode with $V_{\max}/V_{\mathrm{ss}} > 10^{3}$, and $\langle \eta_R \rangle$ represents the mean value averaged over all events in each simulation.}
    \label{fig:phase_dagram_sbmodel_sliplaw}
\end{figure}

\begin{figure*}
    \centering
    \includegraphics[width=\textwidth]{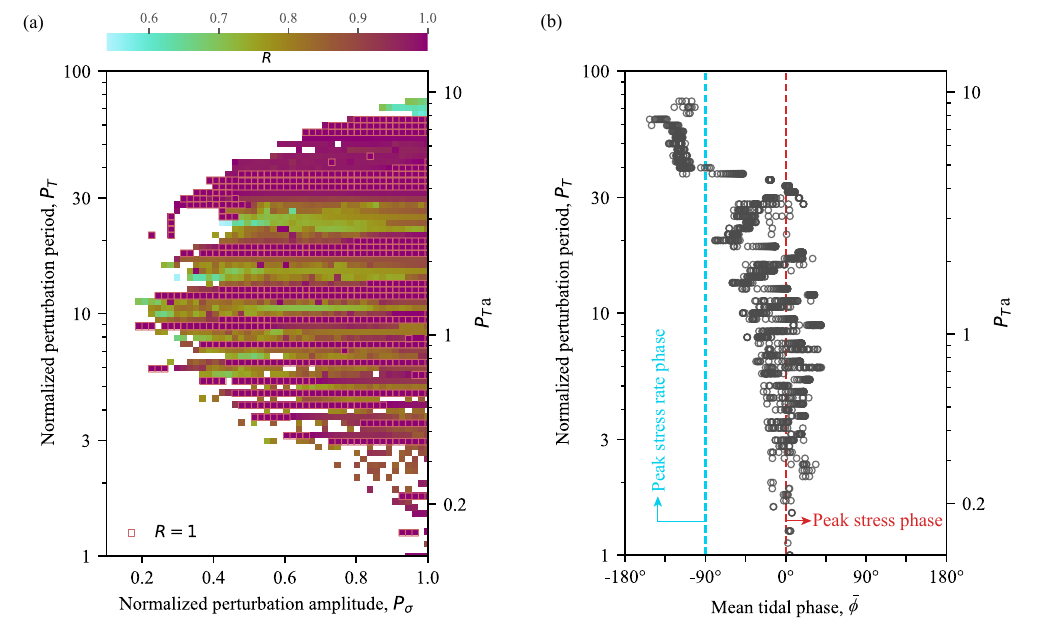}
    \caption{
   Correlation distribution of triggered events for slip law. (a) Color coded by $R$, indicating the dispersion of phase values; the closer $R$ is to 1, the higher the concentration around the mean phase. (b) The horizontal axis represents the mean phase, and the vertical axis denotes $P_T$. }
    \label{fig:R_meanphase_sliplaw}
\end{figure*}

\section{Effect of intrinsic nondimensional parameters}
\label{app:N_kkc_Rab}
\begin{figure}[t]
  \centering
  \includegraphics[width=0.45\textwidth]{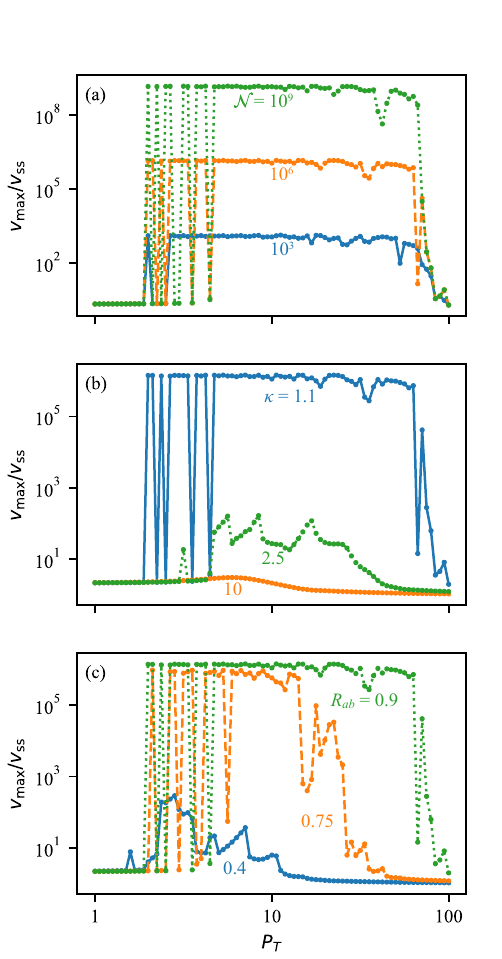}
  \caption{
  Dependence of the normalized maximum slip velocity $V_{\max}/V_{\mathrm{ss}}$ on the normalized perturbation period $P_T$ under harmonic stress perturbations for
(a) different values of $\mathcal{N}$,
(b) different values of $\kappa$,
and (c) different values of $R_{ab}$.
In each panel, only one parameter is varied, while all other parameters are fixed to the reference values listed in Table~\ref{tab:non_dimensional_parameters}.
  }
  \label{fig:n_kappa_rab}
\end{figure}

In addition to the primary control parameters associated with external stress perturbations, we further investigate the influence of the \rev{intrinsic nondimensional system parameters that characterize the internal fault dynamics independent of the external perturbation}, including the quasi-dynamic radiation damping $\mathcal N$, the nondimensional stiffness $\kappa$, and the RSF frictional ratio $R_{ab}$. These analyses aim to assess the robustness of the main results with respect to these three intrinsic system parameters.

Three values of the radiation damping parameter, $\mathcal N = 10^{3}$, $10^{6}$, and $10^{9}$ are tested, which may be interpreted as representing background slip velocity ranging from slow slip velocitys, through typical tectonic plate convergence rates, to an almost stable sliding regime. As shown in Fig.~\ref{fig:n_kappa_rab}a, $V_{\max}/V_{\mathrm{ss}}$ increases systematically with $\mathcal{N}$. Although the scaling is not strictly linear, the overall trend suggests a strong positive dependence of $V_{\max}/V_{\mathrm{ss}}$ on $\mathcal{N}$.
The effect of system stiffness is examined by varying $\kappa = 1.1$, $2.5$, and $10$. As illustrated in Fig.~\ref{fig:n_kappa_rab}b, larger values of $\kappa$ systematically suppress the velocity amplification, indicating a stabilizing influence of increased stiffness.
The friction ratio is varied between $R_{ab} = 0.4$ and $0.75$. 
As shown in Fig.~\ref{fig:n_kappa_rab}c, increasing $R_{ab}$ toward unity enhances the slip velocity amplification, indicating that as the system approaches the velocity-neutral transitional regime $(a \approx b)$ within the VW domain, the effective frictional resistance to perturbations is reduced, making the fault more susceptible to large slip velocity amplifications.

\end{document}